\documentclass[prd,showpacs,superscriptaddress]{revtex4}
\usepackage{epsfig}

\begin{document}

\title{Proton and Neutrino Extragalactic Astronomy}

\author{Paolo Lipari}
\email{paolo.lipari@roma1.infn.it}
\affiliation{INFN  sez. Roma ``La Sapienza'' \\
 Dipartimento di Fisica,  Universit\`a di Roma I, 
P.A.Moro 2, 00185 Roma, Italy}

\begin{abstract}
The study of  extragalactic sources of high energy radiation 
via the direct measurement of  the  proton and neutrino fluxes
that they are likely to emit is one of the main goals 
for  the   future observations   of  the recently developed 
air showers detectors and neutrino telescopes.
In this work we discuss the relation between the
inclusive proton and neutrino signals from the ensemble of all sources
in the universe, and the ``resolved'' signals from 
the closest and  brightest  objects.
We also  compare the sensitivities  of 
proton and neutrino  telescopes  and comment on the relation
between these two new astronomies.
\end{abstract}


% 95.85.Ry   Neutrino, muon, pion, and other elementary particles; cosmic rays
% 96.40.Tv   Neutrinos and muons
% 96.50.Vg Energetic particles 
% 98.70.Sa Cosmic rays (including sources, origin, acceleration, and interactions) 
% 13.35.Hb   Decays of heavy neutrinos
% 14.60.Pq   Neutrino mass and mixing

\maketitle

\section{Introduction}
There is a general consensus that   the highest energy cosmic  rays  (CR)
are  of  extragalactic  origin   because  they reach the Earth  with an approximately
isotropic  angular distribution.  The magnetic
fields of the Milky Way are not sufficiently
strong and extended to  randomize the directions of 
particles  produced   by  our own Galaxy,  and the   this isotropy   of these CR 
is likely to  reflect the  large scale    homogeneity of the universe.
It is  natural to expect that the  extragalactic    Ultra High Energy   Cosmic Rays
(UHECR) are produced in  ``point--like'' astrophysical  sources.
The identification of these  sources, and the clarification of the  mechanisms
that  accelerate  particles to these  very large energies  is clearly one
of the crucial goals of  the   new large acceptance  air shower  detectors like the
Pierre Auger Observatory.  

The  trajectories  of  charged  particles is bent by the presence
of astrophysical  magnetic  fields, however  at sufficiently  large
rigidity $E/Z$ (with $Z$ the  particle electric  charge)  the   
magnetic deviations   should 
become sufficiently   small  to allow the direct  imaging of sources with cosmic rays.
Unfortunately  the  intensity and  structure of   the intergalactic  magnetic  field
are  very poorly known, and therefore   the   region  (in source distance
and particle energy)  where  CR astronomy is possible remains  very uncertain.  
It is possible (and there  are  observational hints)  that  the CR   
above  $E \gtrsim {\rm few} \times 10^{19}$~eV are already propagating in
quasi--linear mode  for  source distances  of  order
100~Mpc or more, however the identification of the sources remains
difficult because of the very small number of events 
available, with  most of the   sources contributing  with not more than a  single  event.

The analysis of the ``clustering'' of the observed events can 
give information about  the luminosity of the individual sources.
After  the  data of AGASA  \cite{Hayashida:1996bc,Takeda:1999sg}    gave  hints of possible clustering.
this question   has received  a  significant amount 
of attention  \cite{Uchihori:1999gu,Dubovsky:2000gv,Sommers:2008ji}.
In this  work  we want to  make  
a more quantitative   and detailed  study on how to interpret the results on the clustering.

The nature  of the  particles   in the UHECR  
remains  a central open questions in the field. In this  work we   will assume
that  most of these particles are protons.
This   hypothesis  is consistent  with  the existing  data  
(if one takes into account the  systematic  uncertainties 
on  the modeling of hadronic showers)  
and it is also  favored  in most theoretical models.
Here this   assumption is also  made to  allow a detailed quantitative
description of particle  propagation in  intergalactic space, which is  essential for
the problem we are considering. 
It is  possible  and  reasonably straightforward to generalize
 the  discussion to the case of nuclei.

All cosmic ray sources are  unavoidably also
sources of neutrinos \cite{Gaisser:1994yf,Learned:2000sw}
 because some   fraction of  a population of
relativistic hadrons  will necessarily 
interact   with   ordinary matter
or radiation fields  targets   inside or near their acceleration site
creating pions and other  weakly decaying particles   that can 
(chain) decay into  neutrinos.
Viceversa,  the emission of neutrinos  imply the 
existence of  relativistic hadrons, and therefore the  acceleration of cosmic rays.
The  catalogues of the CR and neutrino sources  are in principle  identical.
In practice   the situation could more  complicated,
because  the  energy range studied
by  the CR  and neutrino  telescopes   differ by several  orders
of magnitude,  and the  relation  between the  neutrino and
cosmic   ray emission  from a source 
 can  vary significantly  depending on the 
structure of the source, and it is 
 certainly possible  to  have  bright  CR sources that emit a relatively
small  amount of neutrinos, and  viceversa.
Nonetheless, it is  reasonable to expect  that the   brighest
extragalactic sources of both  CR  and neutrinos  will coincide, and it is interesting
to discuss in parallel these new  astronomies.

Also  in the case of neutrinos  one  can  measure
an  inclusive  flux, that sums  the  contributions from all
sources in the universe,   while the 
 brightest  and most powerful  neutrino sources   should  be identifiable as  individual 
objects.  Therefore also  for neutrinos 
 the ``clustering'' of the detected events is  a useful  method of study.
An important  advantage is that   for neutrinos 
linear propagation  is   certain,  but one has to deal with the 
existence  of   the foreground of atmospheric  neutrinos.
The  theoretical  framework   needed to discuss the clustering
of neutrino  events is  essentially identical  to
 the  one  needed for  protons. A  quantitative difference
is  that for  neutrinos  extragalactic  space is perfectly  transparent,
and the  inclusive flux  receives most of its contribution
from  very distant and very faint sources, and  the fraction of
this  flux that can be ``resolved'' in     the contribution of
identified  sources is likely to be   small.
More that the interpretation of the  existing data, the goal of this  work is the development
of some general analysis instruments  that can be used in  the study 
of future, higher statistics results.

This  work is organized as follows:
in the next  section we   make a preliminary discussion 
of the so called ``Olbers Paradox''. The discussion of
this  celebrated puzzle allows to introduce  the  key concepts 
needed  to compute the signal from the  ensemble of all sources in  the universe.
In section~3  we  collect the results on particle propagation
in intergalactic space that are needed in the following.  
Section~4 discusses   how the flux   received   from  one astrophysical source
depends on  its  redshift.
Section~5  discusses   possible forms of the luminosity function
of the proton and neutrino sources.
Section~6 compute the inclusive  particle flux from the
combined  emission  of all sources in  the universe.
Section~7 discusses  what  part of the inclusive flux
can  be  resolved in the contribution  of individual sources.
Section~8  applies these  analysis  instruments 
to the  interpretation of the recent Auger data.
Section~9  discusses  extragalactic  neutrino astronomy.
Section~10  gives  some conclusions.

\section{The ``Kepler--Olbers Paradox''}
In  1610  Kepler  was the first to   note that the darkness
of the  night  sky seems to imply that  the  universe
is   finite, or at least contains  a finite  number of stars.
This    point   was  later  rediscussed by several authors,
and is  commonly  known in the   formulation of Henrich Olbers in 1823.
The   so called  ``Olbers paradox''  is the  statement   that
in an  infinite,   homogeneous and static  universe 
the  surface of the   celestial sphere  should  be infinitely bright.
A  derivation of this  result is  elementary.
In euclidean space,  the   relation between the
apparent  ($\ell$)  and  absolute ($L$) luminosity of a source
placed at a distance  $R$ (neglecting photon  absorption) is:
\begin{equation}
\ell  = \frac{L}{4 \, \pi \, R^2}
\label{eq:f0a}
\end{equation}
An homogeneous   ensemble   of  identical  sources
of  density $n_s$    produces  then an 
inclusive  energy flux   per unit of solid  angle $dF/d\Omega$
that diverges  linearly because  every spherical  layer 
contained  in  the interval   [$R,R+dR]$  
gives  an equal    contribution:
\begin{equation}
\left . 
\frac{dF}{d\Omega} \right|_{\rm inclusive} =  
\frac{1}{4 \, \pi} ~\int_0^\infty  dR~  (4 \, \pi \, n_s \;   R^2)
 ~ \frac {L}{4 \, \pi \, R^2} =
\frac{n_s\; L}{4 \, \pi} ~\int_0^\infty  dR~1 \to \infty
 \label{eq:m0a}
 \end{equation}

Assuming  the same homogeneous distribution
of identical  sources,  
one can also consider 
the  number  of  objects   in the sky  that have  apparent luminosity
in the interval  $[\ell, \ell + d\ell]$.
The  distribution  $dN_s/d\ell$   can be   easily calculated as:
 \begin{equation}
 \frac{dN_s}{d\ell } = \frac{dN_s}{dR} \; \frac{dR}{d\ell} = 
 4 \, \pi ~n_s ~\left [
 R^2~
 \left |
 \frac{d\ell}{dR}
 \right |^{-1}
 \right ]_{R=\sqrt{ \frac{L}{4 \, \pi} }}
 =  4 \, \pi ~ n_s~ \left ( \frac{L}{4 \, \pi} \right )^{3/2} 
 ~ \frac {\ell^{-5/2}}{2} ~,
 \label{eq:g0a}
 \end{equation}
a  result  that is also often stated  in  its integral form:
 \begin{equation}
 N_s(\ge \ell_{\rm min})  =
\int_{\rm \ell_{\rm min}}^\infty d\ell ~ \frac{dN_s}{d\ell }
= n_s~\frac{4 \pi}{3} ~
 \left ( \frac{L}{4 \, \pi \; \ell_{\rm min}} \right )^{3/2} 
%= n_s~\frac{4 \pi}{3} ~
% R_{\ell_{\rm min}}^3 
 \end{equation}
 that has the   obvious geometrical   meaning of  the number of  sources
 contained in a sphere of radius $R_{\ell_{\rm min}} = \sqrt{L/(4 \pi \ell_{\rm min})}$.
The  infinite  brightness of the sky is  due to the
``exploding'' number of very faint sources.
In fact for $\ell_{\rm min}  \to 0$
 the   number  of   observed  sources  diverges $\propto \ell_{\rm min}^{-3/2}$, 
and correspondingly  the   integrated energy flux  
diverges as $\propto \ell_{\rm min}^{-1/2}$.

 These  elementary   results  can be  also   restated  in
 an  adimensional  form, and in terms of
 particle  fluxes   instead of  energy fluxes.
 We  consider identical  sources
 emitting $Q$ particles  per unit time,
 and choose   a reference length  $R_0$
(here the choice of $R_0$ is  completely arbitrary,  in the following
it will be natural to identify $R_0$ with the Hubble length).
 The particle flux  received  from a source
 placed at the distance $R_0$ is:
\begin{equation}
\Phi_0 = \frac{Q} {4 \, \pi \; R_0^2 }
= \frac{L} {4 \, \pi \; \langle E \rangle~ R_0^2}
\end{equation}
Using the notation:
$R = R_0 \; z$ 
and 
$\Phi  = \Phi_0  \; x$  
one can  rewrite  equations
(\ref{eq:f0a}) and (\ref{eq:g0a})   in the equivalent adimensional  form:
\begin{equation}
x \equiv \frac{\Phi}{\Phi_0}   = F(z) = \frac{1}{z^2}
\label{eq:f0b}
\end{equation}
\begin{equation}
\frac{dN_s}{dx} = 4 \, \pi \; \left ( n_s\, R_0^3 \right )
~G(x) = 
 4 \, \pi \; \left ( n_s\, R_0^3 \right )
~\frac{x^{-5/2}}{2} 
\label{eq:g0b}
\end{equation}
In equations  (\ref{eq:f0b})  and (\ref{eq:g0b})  we have
introduced the   two functions 
$F(z)$ and $G(x)$. 
The function   $F(z) =  z^{-2}$
can be called the ``dimming function'', and describes
how the flux from  a  source  decreases as its distance 
increases. 
The  function  
 $G(x)  =  x^{-5/2}/2$
% introduced in equation (\ref{eq:g0a})  describes
can  be called the  ``flux distribution  function''
and    describes how  the number of observable  sources
depends on the size of their  flux.

Summing over all  sources,  the inclusive
flux  per unit solid angle  (as  well as the inclusive energy flux)  diverges;
however  it still formally possible to write  an expression for the
inclusive flux as  an integral over  the scaled  distance $z$:
\begin{eqnarray}
%\frac{d\Phi_{\rm inclusive}} {d\Omega} 
% & = &  \left ( n_s \, \Phi_0  \, R_0^3 \right ) ~ \zeta_\phi
% = \frac{ n_s \, L  \, R_0}{ 4 \, \pi \; \langle E \rangle } ~ \zeta_\phi
% \nonumber \\
%& ~ & \nonumber \\
%
\Phi_{\rm incl} 
& = &  
\frac{R_0}{4 \, \pi} ~ \left ( n_s \, Q \right ) 
~ \int_0^\infty  dz ~K(z) 
= \frac{R_0}{4 \, \pi} ~ \left ( n_s \, Q \right ) 
~ \int_0^\infty  dz ~1 
\to \infty
\label{eq:m0b}
\end{eqnarray}
The function  $K(z) = 1$  can
be called  the  ``Kepler--Olbers   function'',
and  gives the contribution of the spherical  shell
between  (scaled)  distance $z$ and $z+dz$ 
to the inclusive flux.
In general there is  an integral  relation between $K(z)$ and $G(x)$:
\begin{equation}
\int_0^z ~dz~K(z)  =  \int_{F(z)}^\infty dx ~x~G(x)
\label{eq:int0}
\end{equation}
that is  easily understood,
noting  that   both sides   of the equation
describe the  particle flux  generated
inside  a sphere   around the observer  of  (scaled) radius  $z$.

It is obvious that the three functions
$F(z)$, $G(x)$ and $K(z)$ are  related to each other
and,   given one of the three,  
the other two can  be
deduced   taking into account the hypothesis of static,
euclidean space.
For example,    if the ``dimming function''   has  the  general  form   $F(z)$ 
the functions $G(x)$ and $K(z)$  can be  calculated as:
\begin{equation}
G(x) = \left [ \left ( \frac{1}{4 \, \pi \; R_0^3} ~\frac{dV}{dz} \right ) ~ \left | 
\frac{dF(z)}{dz} \right |^{-1} \right ]_{x = F^{-1}(z)}
\label{eq:gg1}
\end{equation}
\begin{equation}
K(z) =  \left ( \frac{1}{4 \, \pi \;R_0^3} ~\frac{dV}{dz} \right ) ~  F(z)
\label{eq:kk1}
\end{equation}
where $dV/dz = 4 \pi \;R_0^3 \, z^2$ is the  volume
contained  in the  spherical  shell $[z,z+dz]$.
In fact equations (\ref{eq:gg1}) and (\ref{eq:kk1})  remain  true  for a realistic  cosmological
model, replacing the volume element $dV/dz$  with the correct general  relativity expression
of equation (\ref{eq:dvol}).

The ``Kepler--Olbers paradox'' is  ``solved'',    
and  we  observe   a   finite   energy  flux   even  if  our universe is  
(to a  very good approximation) euclidean,
(nearly  certainly) infinitely large  and   homogeneous, 
because  the  additional  condition 
of a  static, unchanging universe is  false. 
The calculation of the inclusive flux  from
the ensemble of all extra--galactic sources  must obviously take
into account the expansion of the universe, and the consequent
(redshift) energy loss of all  particles. In addition 
the particles  can  lose energy also for other mechanism
that must be taken into account.
With the inclusion of the cosmological and energy loss effects
the functions $F(z)$, $G(x)$ and $K(z)$  take  
more complex  forms that are different  the  simple result:
\begin{equation}
F(z) = z^{-2}, 
~~~~~~~~~~~~~~~~
G(x) = \frac{x^{-5/2}}{2}
~~~~~~~~~~~~~~~~
K(z) = 1
\label{eq:sol0}
\end{equation}
that  is  valid  for  a 
static, euclidean space universe, and in the absence of  energy losses.
The   form of the functions  $F(z)$, $G(x)$ and $K(z)$   in 
a  realistic  cosmological model
depend on the particle type, 
the particle  energy  and 
encode also the shape  of the emission spectrum
(and clearly also the parameters  that control
the  expansion of the universe),   however
the  solution (\ref{eq:sol0})  will always  emerge as
the universal asymptotic form  
for  small redshift  ($z\to 0$)  and
the largest fluxes  ($x \to \infty$). 
The modifications  have the consequence that  the
inclusive particle flux, obtained integrating  over all possible sources,  
remains  finite.
The  relation (\ref{eq:int0}) 
remains valid  also in the limit  of $z\to \infty$
(and  corresponding  $x\to 0$):
\begin{equation}
\int_0^\infty ~K(z)  =  \int_0^\infty dx ~x~G(x) = \zeta_\Phi
\label{eq:int1}
\end{equation}
The result of the integration  in (\ref{eq:int1}) 
is  finite  and equal to $\zeta_\Phi$.
The  physical  meaning
of $\zeta_\Phi$  can be  understood  qualitatively as the 
equivalent linear size (in units of the Hubble scale length $R_0$)  
of the universe  that  contributes to the flux $\Phi$,
or equivalently as the effective number of Hubble times ($t_0 = R_0/c$)
that   contribute to the flux injection.

\section{Particle Propagation}

\subsection{Energy Losses}
In this  work  we will describe
the energy losses of  neutrinos  and protons propagating in  extra--galactic space 
as a continuous deterministic  process  described by the   function $dE/dt(E,z)$
that   depends  on the particle  energy and  on 
the epoch $z$.  
The  assumption of  continuous energy loss is  
essentially exact for  neutrinos, 
that  lose energy  only because of the  redshift
effect, but is also a  good approximation for protons.
Integrating the   energy loss
over  time  it is possible to compute the 
evolution with time  (or redshift) of  the energy of a particle
constructing  the function  $E_{\rm  evolution} (E_i,z_i, z_f)$ 
that   returns  the  energy  at redshift $z_f$ of a particle
with  energy $E_i$ at   redshift $z_i$:
\begin{equation}
E_f  =  E_{\rm  evolution} (E_i, z_i, z_f)
\label{eq:e_evol}
\end{equation}
Neutrinos, to a very good approximation, only
lose energy for the redshift effect,
and  their energy evolution is:
\begin{equation}
E_f = E_i ~ \left ( \frac{1+z_i}{1+z_f} \right )
\end{equation}
It is particularly important to know the energy at a past
epoch $z$  of a   particle   that is  observed
 ``now''  with energy $E$.
This is  a particular case of the  
general  relation (\ref{eq:e_evol}) and can be indicated 
(propagating backward in time)   as:
\begin{equation}
E_g (E,z) = E_{\rm evolution}(E,0,z)
\end{equation}
For neutrinos one   obviously has:
\begin{equation}
E_g (E,z) = E ~(1+z) 
\end{equation}

\begin{figure} [hbt]
\centerline{\psfig{figure=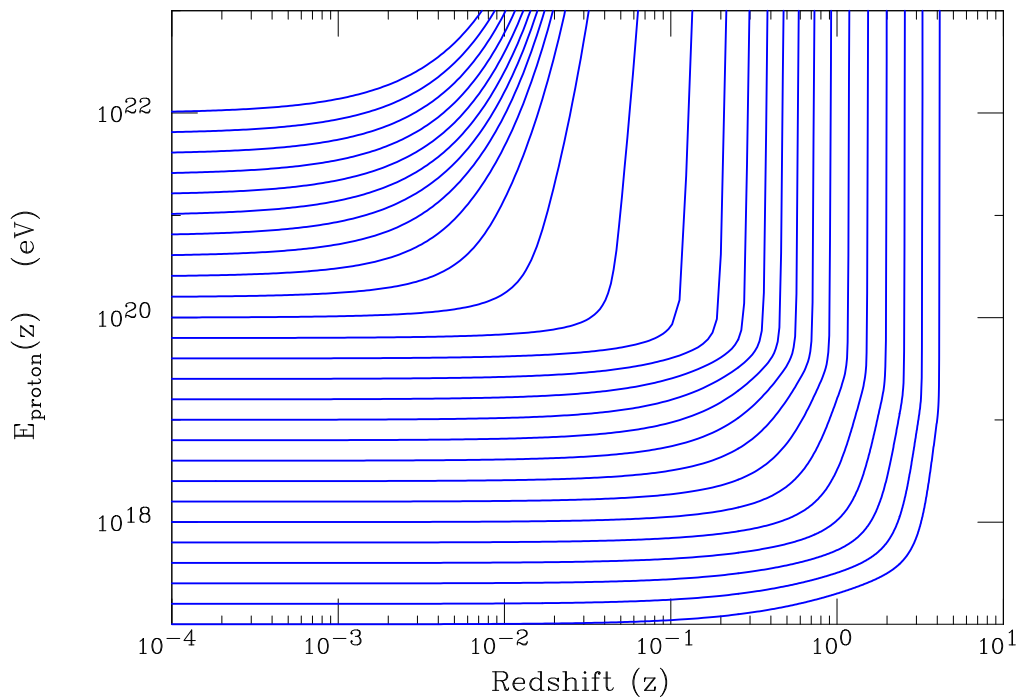,angle=90,width=13.5cm}}
\caption {\footnotesize 
Redshift  evolution of the proton  energy.
The different energies correspond
to protons  observed  at $z=0$ with    energy
$E_{\rm now} = 10^{22- 0.1 , k}$~eV  (with $k=0,$1,$\ldots$,25).
\label{fig:e_evol}  }
\end{figure}

The propagation of protons in  intergalactic  space
has been  discussed in several  works,  after the pioneering work
of Greisen, Zatsepin  and Kuzmin \cite{GZK} 
and is  now well understood.  
Together  with the redshift losses one  has to  include 
losses due to interactions  with the 
photons of the  cosmic microwave background radiation
(CMBR).  The   two    important processes 
are pair production   interactions
($p\gamma \to p \, e^+e^-$)  and
inelastic hadronic interactions (with the threshold processes
$p\gamma \to p \pi^\circ$ and $p\gamma \to n\pi^+$).
The results  of a  numerical calculation  
of $E_g (E,z)$    for protons  are  shown in fig.~\ref{fig:e_evol}.
The different  curves   correspond to the 
``backward in time'' evolution  of protons  detected
``now'' with $E$
between $10^{17}$~eV  and $10^{22}$~eV.
Note that the   time evolution of the proton energy
is   completely independent  from the 
structure and  intensity of the  magnetic  fields,  because  
these fields can  only  bend the trajectory of charged particles, and the
target  radiation field  is  isotropic. 
Obviously, if the particles
do not travel in straight lines, the redshift $z$  
has no simple  relation with the  distance
of the particle source, but only
describes the time of the particle emission.

Inspection of figure~\ref{fig:e_evol}
shows  that  for  each  observed energy $E$ 
there is a critical redshift  where 
the   energy evolution $E(z)$   ``explodes'',
starting to grow  faster than an exponential.
At this  critical redshift the  universe,
filled now with a  hotter and denser radiation field,
has  become  opaque  to the propagation  of the  protons.

\begin{figure} [hbt]
\centerline{\psfig{figure=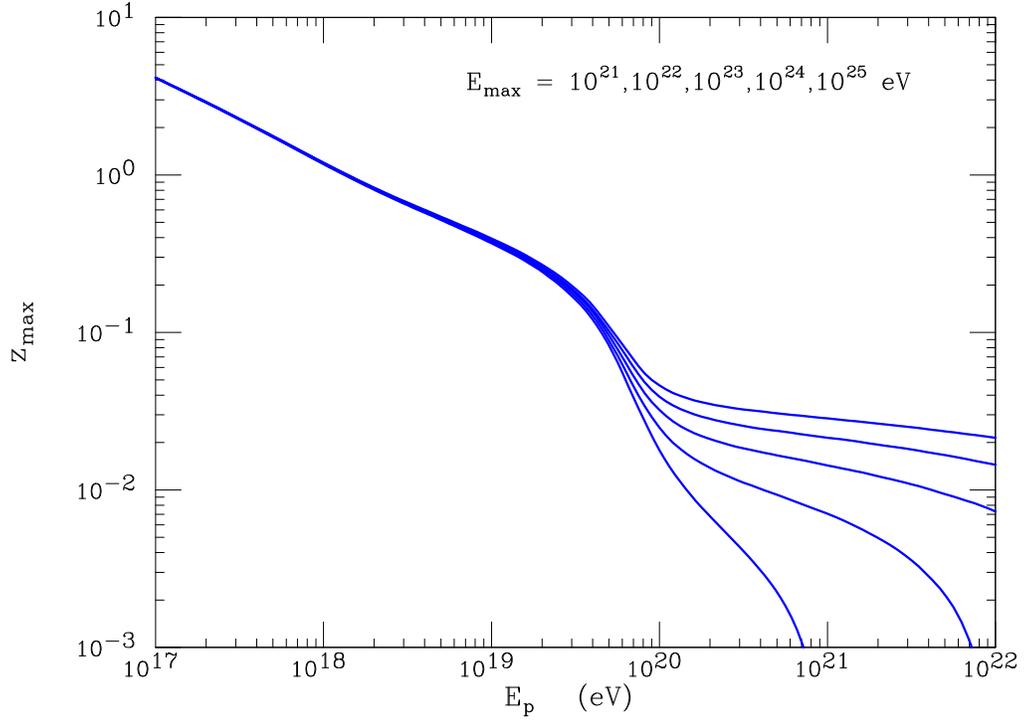,angle=90,width=13.5cm}}
\caption {\footnotesize 
Redshift  horizon  as a function of the
observed proton energy $E$.  The different
curves  are calculated    using
equation (\protect\ref{eq:zh0})   for different  values of  the
maximum   acceleration   energy  $E_{\rm max}$.
\label{fig:zhor}  }
\end{figure}

The form of the redshift evolution of the proton energy suggests 
the introduction of   the concept of  a  redshift ``horizon''. 
The  implicit equation:
\begin{equation}
E_g(E,z_h) =  E_{\rm max}
\label{eq:zh0}
\end{equation}
(with $E_{\rm max}$ the maximum   energy   with which protons are produced in their sources)
can be used as a definition of   the   redshift horizon  $z_h (E,E_{\rm max})$.
An illustration of  this definition of the
 redshift  horizon is  shown in fig.~\ref{fig:zhor}.
The  redshift horizon 
becomes progressively smaller when the   energy   $E$ grows.  When $E$  becomes   a few times
10$^{19}$~eV the  ``shrinking''  of the horizon  with
increasing $E$ accelerates  sharply.  At this   
``GZK''  energy  the protons   start interacting with the
photons of the background radiation in the
{\em  present} universe, and not only with the higher  energy
photons of a  hotter  past.
Above  this  critical  ``GZK energy''  the 
definition of the   horizon  with  equation (\ref{eq:zh0})
strongly depends on the ``maximum acceleration energy''
(that is not a priori a well defined  concept) and therefore
loses  most of  its  usefulness.  A  better  definition 
of the   proton horizon will be given later in equation (\ref{eq:zhor}).

\subsection{Magnetic Deviation}
Protons are charged  particles,  and therefore  their  trajectories
are  bent by  magnetic fields.  Astronomy with  protons is therefore
possible only in a limited  region of   source  distance and  
particle energy    so that  the  observed deviations  are acceptably small.
The size of this  region  is determined    
by the strength and structure the 
astrophysical magnetic  fields.

Neglecting the deviations  due to  to  magnetic  fields
in the vicinity of the source and in the Milky Way halo,
the deviation of a charged  particle 
traveling in intergalactic space for a distance  $d$
can  be   described (in  the limit of small deviation)  with the expression
 \begin{equation}
 (\delta \theta)^2_{\rm extra}   \simeq
 N_{\rm domains} ~\left ( \frac{\lambda_B}{r_L} \right )^2 
=
\left (\frac{d} {\lambda_B} \right )
~\left ( \frac{\lambda_B}{r_L} \right )^2 
 =   ~ (Z \,e \, B)^2 \; \lambda_B~ \left (\frac{d } {E^2} \right )
 \label{eq:mag_dev}
 \end{equation}
where  $r_L = E/(Z \, e \, B)$ is the Larmor  radius,
$B$ is the typical value
 of the extra--galactic magnetic  field and $\lambda_B$ is its coherence
 length. In (\ref{eq:mag_dev})
the magnetic  deviation is modeled  as the  incoherent
sum of contributions  received in  ``magnetic   domains'', where the 
field  has  different  orientation.
In this  simple model, the  angular deviation  
accumulated by a  proton  of  energy $E$
from a  source  at distance $d$   scales   $\propto \sqrt{d}/E$, 
with an absolute  value    controled
by the combination $B\, \sqrt{\lambda_B}$.

If we require  an  image   with an angular  resolution
better that $\delta \theta$,  for  a given  energy $E$  we  must
limit our  observations to   distances smaller
than an  ``imaging radius'':
\begin{equation}
R_{\rm imaging} 
(E, \delta \theta) =  
\frac{ E^2 \, \delta\theta^2}{(Z \, e \, B)^2 \; \lambda_B }
\label{eq:rimaging}
\end{equation}
Alternatively    to produce  an image  with resolution
$\delta \theta$  of a source at a distance
$d$ one  must select a  threshold  energy  $E_{\rm imaging}$
\begin{equation}
E_{\rm imaging} 
(d, \delta \theta) =  
\frac{ \sqrt{d}}{\delta \theta} ~{(Z \, e \, B) \; \sqrt{\lambda_B} }
\label{eq:eimaging}
\end{equation}
 
The  combination $B \sqrt{\lambda_B}$     is therefore  the crucial parameter
that controls  the perspectives  of extragalactic  cosmic ray  astronomy.
The best  (and perhaps  today  the only)
 way to  estimate $B^2 \, \lambda_B$  and
determine where  CR astronomy is possible
is   in fact the   observation of  UHECR
and  the use of equation (\ref{eq:mag_dev}).
At present  however the  
estimate  of $B^2 \, \lambda_B$  remains  very poorly known.

The recent   results of the Auger  collaboration  
\cite{Cronin:2007zz,Abraham:2007si}
indicates that the   directions of the  highest  energy events 
($E \ge 5.6 \times 10^{19}$~eV)  are correlated  
with  the positions of close  AGN's, selected
with  a redhift cut $z < 0.018$  that corresponds to 
$d \simeq 75$~Mpc,   using a   correlation cone of
  opening  angle $\psi \simeq 3.1^\circ$. 
The  simplest  
(``nominal'')   interpretation of these results is
that the  highest energy CR  are in fact  produced in 
these AGN  (or  in   sources very strongly correlated in position)
and  arrive with   a magnetic  deviation of less than 3.1$^\circ$.
This  is  a  remarkable result that, 
attributing all of the observed  deviation to  the intergalactic  propagation
(that is neglecting the contributions of the source
envelope and of the Milky Way), allows to 
set an upper limit on  the combination  $B^2 \, \lambda_B$ 
that describes the extragalactic  magnetic  field:
\begin{equation}
Z^2 \,B^2 \; \lambda_B \le 0.15 ~({\rm nGauss})^2~{\rm Mpc}
\label{eq:estim1}
\end{equation}
Assuming that  the particles  are protons, this  implies 
the imaging radius:
\begin{equation}
R_{\rm imaging}  =  
210~
\left (
\frac{ E}{10^{20}~{\rm eV}} 
\right )^2
~
\left (
\frac{\delta \theta_{\rm image} }{3^\circ} 
\right )^2
~{\rm Mpc}
\label{eq:r_imaginga}
\end{equation}
This  result, if confirmed,  indicates   that  CR astronomy
should be able to  give important  results in the  very near  future.

The  ``nominal interpretation'' of the Auger results,
suffers from  several difficulties 
both internally (see  for example the criticism in \cite{Gorbunov:2008ef})
and comparing with the  observations  of other  detectors
(for the Yakutsk data see
\cite{Ivanov:2008it}  for the HiRes data see \cite{Abbasi:2008md}).
The consensus is that  the  Auger   results 
do indicate the existence of an 
anisotropy of the arrival directions  of the highest energy
cosmic rays,  but the  origin of  the anisotropy has not
been established, the  significance of  the 3.1  degrees of the 
opening of the correlation cone is therefore  ambiguous and the estimates
(\ref{eq:estim1}) and (\ref{eq:r_imaginga}) only tentative.

A  remarkable    fact of the  Auger results is that
two  of the highest energy events  are in correlation
(using the 3.1$^\circ$ cone) with  the nearby 
($d \simeq 3.5$~pc) AGN  Cen~A.
Taking into account the  fact that Cen~A  has  been  often
discussed  as one of the most likely sources of UHECR 
this is a striking    result, that  suggests  that  the first 
extragalactic  CR source has  in  fact already been imaged.
%This result   is certainly  reason for  celebration.
There are   also indications
that  Cen~A could be    the source 
of a larger  number of  events   with a larger  angular spread
(see for example  the remarks in 
\cite{Wibig:2007pf,Gorbunov:2008ef,Stanev:2008sd}).
For example a  20 degrees  cone around Cen~A  contains
9~events.  Such a   large  contribution from Cen~A     could account
for a   significant  fraction of the observed  anisotropy effect.
These  observations  could however  also a  dramatic  impact
on the  estimate of the extragalactic  magnetic  field,  because
they suggest a  larger deviation ($\delta \theta \sim 10^\circ$) for
a   shorter distance.  The  estimate  for the combination
$B^2 \; \lambda_B$    becomes  therefore larger  
by a  factor of order 200:
\begin{equation}
Z^2 \; B^2 \; \lambda_B \simeq 40 ~({\rm nGauss})^2~{\rm Mpc}
\label{eq:estim2}
\end{equation}
with  a correspondingly smaller  imaging radius, and significantly
restricting the potential of CR astronomy for
more  distant sources.
Additional  data should  allow to  clarify  this
ambiguous  situation.
The  discrimination  between  the  two  estimates
(\ref{eq:estim1})  and
(\ref{eq:estim2})  for the  parameter  that describes the 
extragalactic magnetic  field is obviously crucial for the future of
the field.

\section{Flux from  an individual source}
Let us consider a source 
that emits  isotropically particles with  the energy 
spectrum described by the function  $q(E)$.
The  emitted  particles  travel, at least in good
approximation,  along geodetic  trajectories losing  energy
continuously. Propagating backward in time, a particle 
that is   observed  ``now''  with energy $E$, at the time
that correspond  to   redshift $z$  had  energy  $E_g(E,z)$.

The  flux received from  such a   source
placed at the distance that  corresponds
to  redshift $z$ can  be written as:
\begin{equation}
 \phi (E, z) = \frac{q (E) } {4 \pi \, R_{0}^2} ~  F_{\phi}(z,E)
\label{eq:phi_gen}
 \end{equation}
where:
\begin{equation}
R_0 = \frac{c}{H_0}
\end{equation}
is the Hubble  length, and 
the  adimensional  function $F_{\phi}(z,E)$  is  given by:
\begin{equation}
 F_{\phi}(z,E) = \frac{1} { r^2 (z) \, (1+z)} 
~ \frac{q[E_g (E,z)]}{q(E)}   ~    \frac{dE_g(E,z)}{dE} ~.
\label{eq:f_gen}
\end{equation}
In equation (\ref{eq:f_gen})  $E_g(E,z)$ is the emission energy 
(at the source) of  the particles observed  with energy $E$,
and  $r(z)$ is the (adimensional) comoving coordinate  that corresponds
 to the redshift $z$:
 \begin{equation}
 r(z)  = \int_0^z  \frac{dz^\prime}{{\cal H} (z^\prime)}
 \label{eq:rcomoving}
 \end{equation}
 with    ${\cal H}(z) = H(z)/H_0$  the  
(scaled) Hubble parameter  at the epoch of redshift $z$:
 \begin{equation}
 {\cal H} (z) = 
 \sqrt { \Omega_{\rm m} \,  (1 +z)^3 + \Omega_\Lambda +
 (1 - \Omega_{\rm m} - \Omega_\Lambda) \, (1+z)^2 }.
 \label{eq:hubble}
 \end{equation}
The  expansion of the universe, and
the redshift dependence of the Hubble constant  are determined  by  the 
parameters $\Omega_{\rm m}$ and  $\Omega_{\Lambda}$. 

The ``dimming function'' $F_{\phi}(z,E)$  describes how the flux
of a  source   weakens, and (in general) has
its energy  spectrum  distorted  when the
redshift $z$ increases.
The  function  is  the generalization of
Gauss law  ($1/z^2$ behaviour)  that is  valid
for static, euclidean space in  the absence of energy loss.
For small $z$   cosmological  and energy loss
effects  become  negligible  and  the function  $F_{\phi}(z,E)$
takes  the simple asymptotic form
 $F_{\phi}(z,E) \to   z^{-2}$. 
The   form of  the function $F_{\phi}(z,E)$ 
is  determined  by the cosmological parameters
($\Omega_{\rm m}$ and $\Omega_{\Lambda}$) and  from the shape  of the
 emission spectrum. In  principle an
unambiguous  notation should include these  additional 
dependences: 
$F_{\phi}(z,E) \to F_{\phi}(z,E; \{q\}, \Omega_{\rm m}, \Omega_\Lambda)$. For simplicity,
in this  work 
we will leave the dependence on the cosmological parameters and the 
shape of the emission function implicit.

A  relation analogous  to   (\ref{eq:phi_gen}) 
can also be written for the integral  spectrum:
\begin{equation}
 \Phi (E_{\rm min}, z) = \frac{Q (E_{\rm min}) } {4 \pi \, R_{0}^2} 
~  F_{\Phi}(z, E_{\rm min})
\label{eq:Phi_gen}
 \end{equation}
where   $Q(E_{\rm min})$ is the integral  source  emission:
\begin{equation}
Q(E_{\rm min}) = 
\int_{E_{\rm min}}^\infty  dE  ~q(E) ~.
\end{equation}
The function
$F_{\Phi}(z,E)$   (for the integral flux)  can be obtained
from $F_{\phi}(z,E)$  (for the differential flux)  with the integration:
\begin{equation}
F_{\Phi}(z,E_{\rm min}) = 
\frac{1} {Q(E_{\rm min})} 
~\int_{\rm E_{\rm min}}^\infty q(E) ~F_{\phi}(z,E)
\label{eq:f_int}
\end{equation}
For  $z\to 0$, the asymptotic  behaviour  of  $F_{\Phi}(z)$
takes  again the simple form $z^{-2}$.

For neutrinos, and more in general in all cases
when the only significant source of energy loss
is the redshift, the energy  evolution of a particle
has the simple form $E_g (E,z) = E \, (1+z)$
the general expression for $F_{\phi}(z,E)$ given in equation
(\ref{eq:f_gen})     can  be simplified,    obtaining:
 \begin{equation}
  F_{\phi}(z,E)  =   \frac{1} { r^2 (z)} 
 ~\frac{q [E(1+z)])} {q(E)}  
\label{eq:f0nu}
 \end{equation}
As a consistency check one can  observe that 
a source of  such  ``non interacting''  particles
of absolute luminosity
\begin{equation}
L = \int_0^\infty dE~E~q(E)
\end{equation}
placed at a distance $z$,  has an apparent luminosity:
 \begin{eqnarray}
\ell & =  &  \int_0^\infty dE ~E~ \phi (E)  =
 \int_0^\infty dE ~E~ \left (\frac{1}{4 \, \pi \, R_0^2} 
~\frac{1}{r^2(z)} ~q[E(1+z)] \right ) \nonumber \\
& ~ &  \nonumber \\
&  =  & \frac{1}{4 \, \pi \, R_0^2 \; r^2(z) \, (1+z)^2}  ~ 
\left [\int_0^\infty  dE_0~E_0~q(E_0)  \right ]
= \frac{L} {4 \, \pi \, d_L (z)^2}
 \end{eqnarray}
This  expression has the well known form
$\ell = L/(4 \pi \, d_L^2)$,   where 
$d_L$ is the    ``luminosity distance'':
\begin{equation}
 d_L(z) = R_0 \; r(z) \, (1+z) ~.
\end{equation}

\begin{figure} [hbt]
\centerline{\psfig{figure=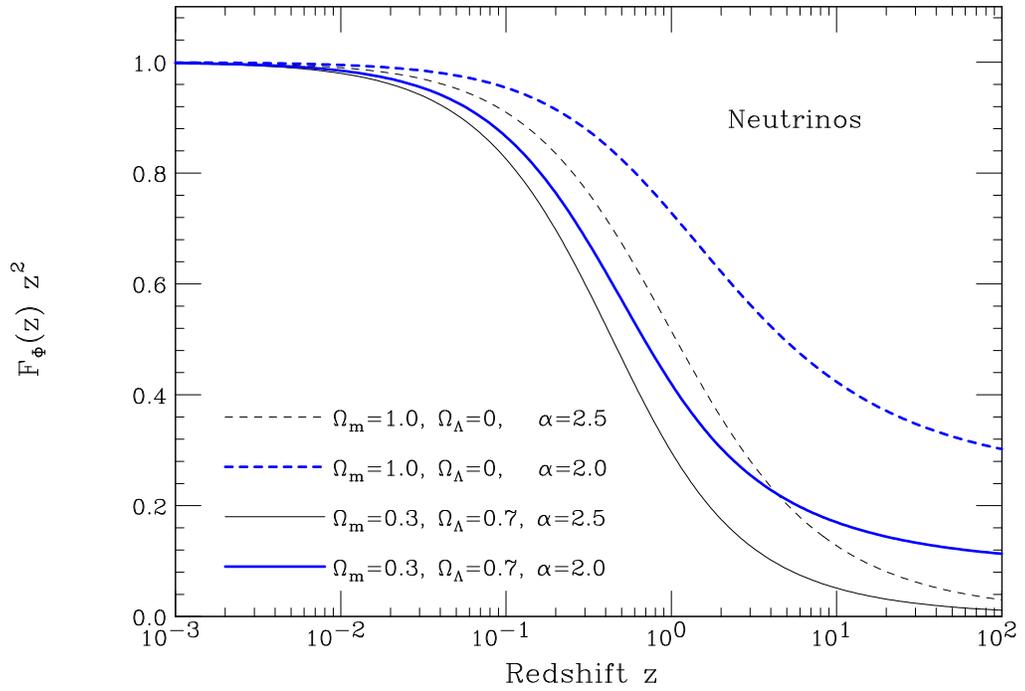,angle=90,width=13.5cm}}
\caption {\footnotesize 
Plot of the function $F_{\phi}(z) = F_\Phi(z)$ that   gives  the
redshift dependence
of the  flux from a  $\nu$ source.
The source spectrum  is  a 
power law of slope $\alpha$. 
The different  curves are calculated  for two different
choices of the cosmological parameters   ($\Omega_{\rm m}$ and
$\Omega_{\Lambda}$),  and  two different
slopes   ($\alpha = 2$  and  $\alpha =  2.5$).
\label{fig:nu1} }
\end{figure}

The ``dimming   functions'' for neutrinos  
take  a very simple form  if one makes the assumption  that the 
emission spectrum   is  un unbroken power law of  slope $\alpha$.
In this case   the functions $F_{\phi}(z)$ and $F_{\Phi}(z)$ 
become  independent from energy,  and equal to each other:
  \begin{equation}
  F_{\phi}(z) =  F_{\Phi}(z) = \frac{(1+z)^{-\alpha}}{r^2(z)}
  \end{equation}
In case of a power law emission  of neutrinos, the spectrum  mantains its (featureless)  shape 
even after propagation and energy loss.
 The  function $F_{\phi}(z)$  for neutrinos is
shown in figure~\ref{fig:nu1}.  The function   depends
  on  the  value of the cosmological parameters
  $\Omega_{\rm m}$ and $\Omega_\Lambda$, and of the  value of the
  slope  $\alpha$.
One can  see that $F_{\phi}(z)$ coincides  with the 
form  $z^{-2}$ for $z \lesssim 0.1$,  and then  begins
to  fall  more  rapidly suppressing the flux  of   very distant
sources.

\begin{figure} [hbt]
\centerline{\psfig{figure=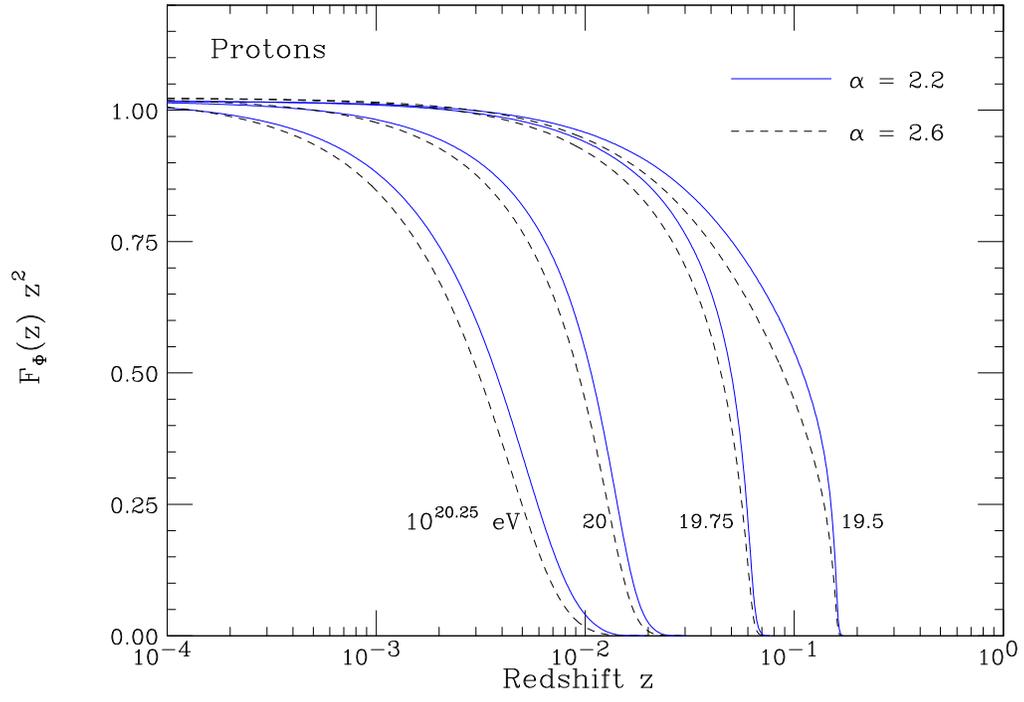,angle=90,width=13.5cm}}
\caption {\footnotesize 
Plot of the function $F_{\Phi}(z,E_{\rm min})$  
(in the form  $z^2 \, F_{\Phi}(z)$  versus $z$)  for protons.
The  function $F_{\Phi}(z)$   gives the
``dimming'' of the integral proton  flux 
above the minimum  energy $E_{\rm min}$ 
with increasing  redshift $z$. The solid  (dashed) curves
are calculated  for a proton spectrum slope 
with slope $\alpha = 2.2$ ($\alpha = 2.6$), and different
values of the threshold energy $E_{\rm min}$.
\label{fig:f_prot}  }
\end{figure}

The  ``dimming function'' for  proton  sources   depends
very  strongly on the proton energy.
Some examples  of this behaviour (for  the integral  flux) are  shown in fig.~\ref{fig:f_prot}
where the   redshift dependence of the function $F_{\Phi}(z,E_{\rm min})$  is shown
for  several  value  of the threshold energy $E_{\rm min}$, assuming
again a power law  emission spectrum.
One can  clearly  see that  a  proton source  becomes   rapidly unobservable
when it becomes  more distant  than a rather  sharply
defined  ``horizon''  that  shrinks  rapidly with energy for 
$E_{\rm min}  \simeq 10^{19}$ to  $10^{20}$~eV,
and is much smaller that the Hubble  length.

\section{Luminosity Function}
In this  work  we will   make  the assumption that 
all astrophysical sources  have emission 
of  the same  shape and differ  only  for their
absolute  luminosity.
The idea  behind this  assumption   is that the emission  mechanism
is  likely to be universal,  independent from the 
properties of the  individual  sources. 
An example of this situation is
the acceleration   of cosmic rays
 near the   blast waves of  SuperNova (SN) explosions,
where it is  believed that  particles are accelerated 
with a  power law spectrum  $\propto E^{-\alpha}$
with a   slope $\alpha  \simeq 2$  that is   essentially 
independent from  the properties of an individual   SN
(for example  the   mass and   initial  velocity of the ejected material),
but it is  determined only by the  (universal) compression  factor
of the interstellar  gas  after the  passage of a  strong  shock front.

This   assumption   of a ``universal'' spectral shape  is 
of course an  hypothesis that could very well be   incorrect.
This possibility  is in fact  likely to 
the UHE  proton sources, 
since it is  in fact natural to expect
that different  sources have  different high energy cutoffs,
and that these cutoffs  are  in the  energy range
where the spectra are measured.
After these  words of  warning, we will anyway  
make the  assumption  of a universal  shape  as it 
clearly  significantly  simplifies the discussion, and represents
a natural   first order  approximation.

The luminosity function $dn_s/dL(L,z)$  gives  the 
number density    (in comoving volume)  of sources
of  luminosity $L$ at   the epoch of redshift $z$.
The power density ${\cal L}(z)$ of the sources at redshift $z$ 
is  given  integrating over  all luminosities:
\begin{equation}
{\cal L}(z)  = \int dL ~L~\frac{dn_s}{dL}(L,z)
\end{equation}

In  this  work   we   will   discuss   emission spectra
that are well   represented (at least 
in the region of  interest of the measurements)
as  simple  power laws of constant slope. 
In this  situation however, the  source  
bolometric luminosity 
is  not  unambiguously  defined 
because the integral 
$$
\int_0^\infty  dE~E~E^{-\alpha}
$$
 is  always divergent
(at the upper limit for $\alpha \le 2$,  and at the lower 
limit  for  $\alpha \ge 2$).
In other words,  an unbroken  power law emission   is impossible,
and to estimate the   bolometric
luminosity of a source it is necessary to specify 
the range of validity of the   power law form, and
the  behaviour  of the  spectrum  outside this  region.
This  requires the introduction 
of additional parameters that are not easily constrained, because 
one  observes only a  limited  range  of the spectrum.
What is  really needed for our  purposes is 
to  determine the  absolute normalization
of the power law  spectrum in  the energy region of interest.
This  can be  done   considering  the luminosity
in a  fixed,  relevant energy  range.
In this  work we will use the convention to  use
the luminosity  (or luminosity density) 
integrated over  one  energy  decade
starting from a minimum energy $E_{\rm min}$  that   is  relevant
for  the observations
\begin{equation}
L \equiv L_{\rm decade} (E_{\rm min}) 
= \int_{E_{\rm min}}^{10 \, E_{\rm min}} dE~q(E)
\end{equation}
(in the case of $\alpha = 2$:
the luminosity per energy decade is   constant and independent
from  $E_{\rm min}$):
%In  practice we will chose $E^* = 10^{12}$~eV   when  discussing neutrinos,
%and $E^* = 10^{19}$~eV  discussing protons
To keep the notation simple  in the following
the subscript ``decade'' will be omitted, and
the dependence  $E_{\rm min}$ (in most cases) will  be left implicit.
The integral emission   of particles
above the   energy $E_{\rm min}$ is  related
to  this  definition of the luminosity by the relation:
 \begin{equation}
 Q(E_{\rm min}, L) = \frac{L ~\eta }{E_{\rm min}}
 \end{equation}
where  $\eta$ is an  adimensional  factor  that depends  on  the shape
of the spectrum.  For a simple  power  law of slope $\alpha$ one has:
\begin{equation}
\eta = \eta_{\alpha}  = \frac{(\alpha -2)}{(\alpha -1)}
\left ( 1 -10^{-(\alpha-2)} \right ) ~.
\label{eq:eta}
\end{equation}
For   $\alpha\to 2$  the  limit of equation (\ref{eq:eta}) is 
$\eta_2 = 1/{\ln 10} \simeq 0.434$. 
The  differential and integral spectra
for an arbitrary value   of the energy 
(and assuming the  validity  of the  power law behaviour):
 \begin{equation}
 q(E,L) =
\left [\frac{ L 
~ \eta_\alpha \; (\alpha -1)}{E_{\rm min}^2}
 \right ] ~
~   \left ( \frac{E}{E_{\rm min}} \right )^{-\alpha} 
 \end{equation}
 \begin{equation}
 Q(E,L) =
\left [\frac{ L ~ \eta_\alpha }{E_{\rm min}}
 \right ]
~   \left ( \frac{E}{E_{\rm min}} \right )^{-\alpha+1} 
 \end{equation}

\begin{figure} [hbt]
\centerline{\psfig{figure=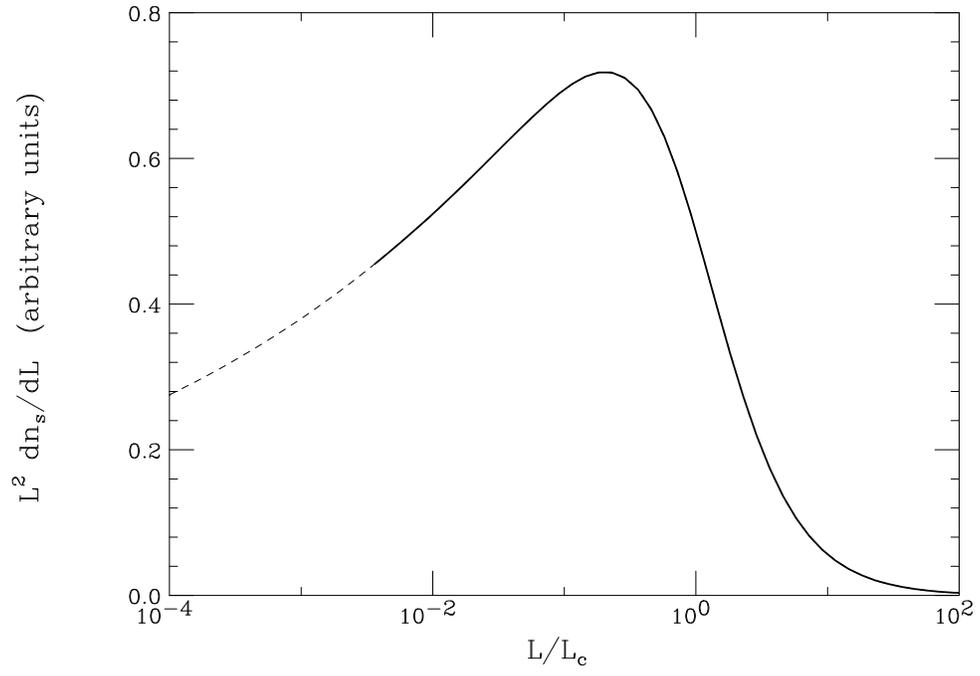,angle=90,width=12.9cm}}
\caption {\footnotesize 
Shape of the luminosity function  that  fits
the   $X$--ray observations  of  AGN  in the  [10,20]~KeV band
(from  \protect\cite{ueda}). 
\label{fig:lum_shape}  }
\end{figure}

The simplest hypothesis  for  the luminosity function
is  to assume  that all sources are identical
with luminosity $L_0$:
\begin{equation}
\frac{dn_s}{dL} = \delta [L - L_0]
\end{equation}
This  is  nearly  certainly    incorrect.
As an  example,  the luminosity  function 
of AGN in the  hard  $X$--ray band  ([10,20]~KeV)
   has  been   fitted by  Ueda \cite{ueda}
with the  functional form:
\begin{equation}
\frac{dn_s}{dL} (L) = \frac{n_0}{L_c} ~
\left (\frac{L_c}{L} \right ) ~
 \left [ 
\left (\frac{L}{L_c} \right )^{0.86}  + 
\left (\frac{L}{L_c} \right )^{2.23} 
\right ]^{-1}
\label{eq:lum_shape}
\end{equation}
with  $L_c \simeq  10^{43.94}$~erg/s. 
This  shape   provides  a good fit  
for all luminosities 
$L_x \in [10^{41.5},10^{46.5}]$~erg/s,   
in a  broad range that extends for  5 orders of  magnitude.
In the following  we will use expression (\ref{eq:lum_shape})
(shown in figure~\ref{fig:lum_shape})
as a ``template''    (with two  free parameters
$n_0$ and $L_c$, that in principle can also
be   redshift  dependent)  to build  examples of 
the  luminosity functions  for  the 
(yet undiscovered)  proton and  neutrino extragalactic sources.

Note that   the integration of the expression (\ref{eq:lum_shape}) 
for  the luminosity function     diverges at the   lower  limit
(corresponding  to a  divergent number of  faint sources)  while the 
 power  density remains  finite.  This fact illustrates the point that the
concept of ``total number of sources'' 
should  be  treated  with  caution,
because the  existence of a large number of     weak sources
is  difficult to  establish  (or refute).
In the  following 
we will  use the  form of the  luminosity distribution 
(\ref{eq:lum_shape})  assuming  also a low
luminosity sharp cutoff:
\begin{equation}
L \ge  3 \times 10^{-3}\; L_c
\label{eq:lum_cutoff}
\end{equation}

In this work  we will  use as the 
``characteristic''   value of the
source  luminosity the median luminosity $L_0$
of the sources at  redshift $z=0$, 
defined as the solution of the   equation
\begin{equation}
%\int_{L_{\rm min}}^{L_0} dL ~L ~\frac{dn_s}{dL} (L,0) =
%\int^{L_{\rm max}}_{L_0} dL ~L ~\frac{dn_s}{dL} (L,0)~.
\int_{0}^{L_0} dL ~L ~\frac{dn_s}{dL} (L,0) =
\int^{\infty}_{L_0} dL ~L ~\frac{dn_s}{dL} (L,0)~.
\label{eq:lmedian}
\end{equation}
Half of the  power density
is due to sources with $L \le L_0$, and the other half
to sources with $L > L_0$.
For the  distribution  (\ref{eq:lum_shape})
with the cutoff (\ref{eq:lum_cutoff}) one has  $L_0 = 0.116~L_c$.

Without loss of generality, it  is possible to write
the luminosity function in the form:
\begin{equation}
\frac{dn_s}{dL} (L,z) =
\frac{{\cal L}(0)}{L_0^2} ~f_L \left ( \frac{L}{L_0}, z\right )
\label{eq:flum}
\end{equation}
with ${\cal L}(0)$ the power  density of the
ensemble of all sources at $z=0$, and 
  the adimensional  function $f_L(y,z)$ 
satisfies the condition:
\begin{equation}
\int_0^\infty  dy~y~f_L(y,z) = \frac{{\cal L}(z)}{{\cal L}(0)}~.
\label{eq:flum2}
\end{equation}

\begin{figure} [hbt]
\centerline{\psfig{figure=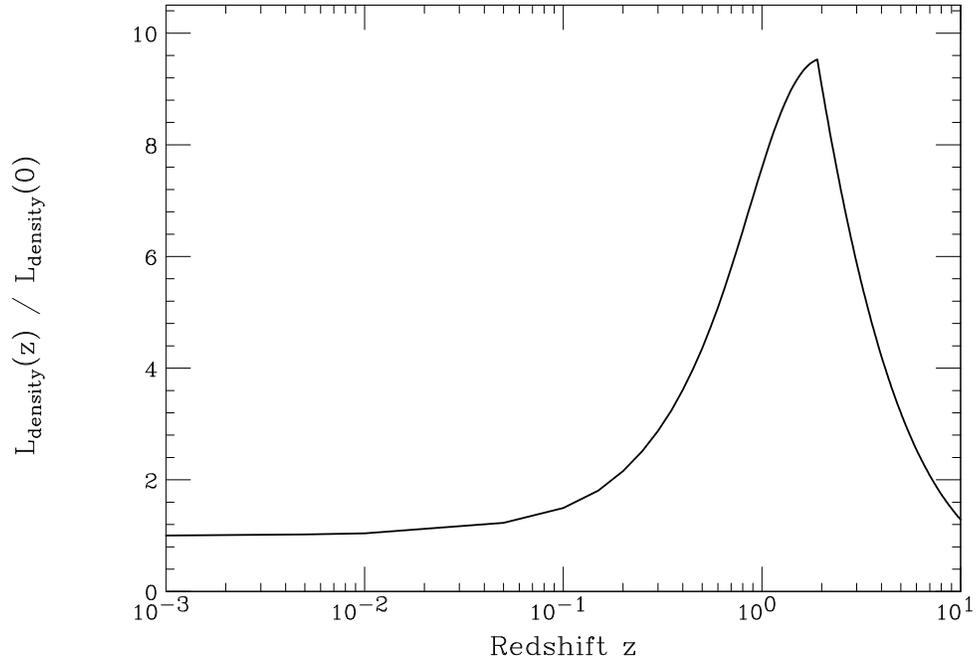,angle=90,width=12.9cm}}
\caption {\footnotesize 
Redshift evolution  of the X--ray luminosity density of AGN
in the [10,20]~KeV band according to the fit 
of \protect\cite{ueda}.
\label{fig:lum_evol}  }
\end{figure}

The redshift dependences of the  power density of the
extragalactic cosmic  ray and neutrino  sources  is of course not known,
however it is  likely that they have   behaviours that
are similar  to each other, and that are also similar
to the evolution with cosmic time of the star formation rate,
or of the AGN activity.  It is  therefore
expected that  the power density ${\cal L}(z)$  
will  grow  with increasing  $z$, before
falling of at very large   redshifts  that correspond to early times
before the formation of the first sources.
As an example  of  possible redshift evolution,
in  figure~\ref{fig:lum_evol}  we
show  the  evolution of the AGN  luminosity in
the   hard X--ray  band ([10,20]~KeV) as fitted
by Ueda \cite{ueda}.

 \section{Inclusive  Flux}
Summing over all sources, it is possible to define
the inclusive particle ``emissivity''  $q_{\rm incl}(E,z)$
that  gives   the 
number of  of particles of energy $E$  emitted per unit time and  unit  of
comoving  volume  at the epoch of redshift $z$.
In  the simplified  model   where all sources  have
emission  of the same  spectral  shape,  and the  ``median source''  
(of luminosity $L_0$) has emission $q(E)$, the emissivity  takes  the 
factorized form:
\begin{equation}
q_{\rm incl} (E,z)  = q(E) ~\int dL~ \frac{dn_s}{dL} (L,z) ~\frac{L}{L_0} 
=
\frac{ {\cal L}(z)} {L_0} ~q (E)
\label{eq:qincl0}
\end{equation}
From the emissity is  is possible to obtain  the space
averaged  particle  density  at the present epoch 
$\langle n(E) \rangle$  as:
\begin{eqnarray} 
\langle n(E) \rangle  & =  &
\int dE_0 ~\int_0^{t_0} dt  ~q_{\rm incl}(E_0,t)~ \delta [E -E_{\rm  evol}(E_0,t,t_0)]
\nonumber \\
& ~ & \nonumber \\
& = & \frac{q_{\rm incl}(E,0)}{H_0} ~
 \int_0^\infty dz~
\frac{1}{{\cal H}(z) \; (1 +z) }
 ~  \frac{q_{\rm incl}[E_g(E,z),z]}{q_{\rm incl}(E,0)}  
 ~ \frac{ dE_g(E,z)}{dE} 
\label{eq:ndens0}
\end{eqnarray}
The  first line in    (\ref{eq:ndens0}) 
simply says  that particles  observed  now with energy
$E$  have been produced in the past  at time  $t$
 with  initial energy $E_0$;  the integration
over all possible  times and energy
is limited by a delta  function because we are 
assuming  that the  evolution
with time of  the particles  energy is  deterministic, and therefore
for  each emission time   $t$,
 only particles with a well defined  $E_0(E,t)$ can 
contribute to the density of particles  with energy $E$.
The result  of equation (\ref{eq:ndens0})   can then  be obtained
transforming the integral over time  in one 
over  redshift using the Jacobian   factor
\begin{equation}
\frac{dt}{dz} = -\frac{1}{H(z) \; (1+z) } = -\frac{1}{H_0} 
~ \frac{1}{{\cal H}(z) \; (1+z)}
\end{equation}
and performing the integration over 
the emission  energy $E_0$ with use of  the delta function.

If we make the assumption that  all  sources have identical
spectral shape, and therefore that the emissivity is  factorized
with the form (\ref{eq:qincl0})  the  we can  recast equation (\ref{eq:ndens0})
in  the form:
\begin{eqnarray} 
\langle n(E) \rangle  & =  &
\frac{q_{\rm incl}(E,0)}{H_0} ~
 \int_0^\infty dz~
\frac{1}{{\cal H}(z) \; (1 +z) }
 ~  \frac{q[E_g(E,z)]}{q(E)}  
 ~\frac{ {\cal L}(z)}{ {\cal L}(0)}
 ~ \frac{ dE_g(E,z)}{dE}
\nonumber \\
& ~ & \nonumber \\
& = & \frac{q_{\rm incl}(E,0)}{H_0} ~
 \int_0^\infty dz~
\frac{r^2(z)}{{\cal H}(z) }
~F_{\phi}(z,E)
~\frac{ {\cal L}(z)}{ {\cal L}(0)}
\label{eq:ndens}
\end{eqnarray}
where we also have used the definition of 
$F_{\phi}(z,E)$ in equation (\ref{eq:f_gen})

It is important to stress the fact that this  result
describes the  {\em space averaged} 
particle density.
The particle  density ``here''  (meaning near the Earth,
or in the disk of the Milky Way)  is  of course
a  directly   measurable  quantity, and  does not
necessarily coincide  with  the space--averaged one.
The ``danger'' is that  intergalactic 
(or source  envelope)  magnetic  fields can  be sufficiently strong
so that the emitted   particles 
can only diffuse  away  slowly   from their sources.
The extragalactic  cosmic rays  could therefore form
slowly expanding bubbles around their sources,
and be homogeneous only  after  averaging over  a distance
scale  larger than the average source  separation.
Since  the diffusion is  a function of  rigidity this
situation could result in a variety of 
different interesting possibilities, depending on  the 
energy dependence of the  diffusion coefficients and the 
position of the  observer with respect to the sources.
If   the particles  during their lifetime have the time
to propagate  for a linear distance comparable to
the source separation  then  
the extragalactic  cosmic  ray gas  becomes
homogeneous,  these complications 
can be safely neglected,  and we can  identify the
density of the observable  
extragalactic cosmic  rays  with  the
space averaged one  given in equation (\ref{eq:ndens}).
This  same  condition  on CR 
propagation also   garantees  the isotropy of the particles,
and  
the  observable inclusive particle  flux
$\phi_{\rm incl}(E)$  (in  units (cm$^{2}$s~sr)$^{-1}$)  
can be estimated as:
\begin{equation}
\phi_{\rm incl} (E) = \frac{c}{4 \, \pi} ~\langle n(E) \rangle
%= \frac{R_0} {4 \, \pi} ~q_{\rm incl}(E) ~\zeta_\phi(E)
\end{equation}
It is  useful to   introduce  notations for  the integral
(and the integrand)  over  the  redshift $z$  
in equation (\ref{eq:ndens}):
\begin{equation}
\int_0^\infty dz~ 
\frac{r^2(z)}{{\cal H}(z) } ~F_\phi(E,z)
~ \frac{{\cal L}(z)}{{\cal L}(0)}
=
\int_0^\infty dz~ K_\phi(z,E)
=
 \zeta_\phi(E) 
 \label{eq:xia}
\end{equation}
With  these   definitions the inclusive  flux  
can be written in   the form:
\begin{equation}
\phi_{\rm incl}(E) = \frac{R_0}{4 \, \pi} ~q_{\rm incl}(E,0) ~\zeta_\phi(E)
\label{eq:phi_incl}
\end{equation}
The function  $K_\phi(z,E)$  in equation (\ref{eq:xia}) 
has the meaning  of the ``redshift response'',
and gives the contribution of  the  time interval   that corresponds 
redshift in $[z,z+dz]$ to the   
particle flux  at energy $E$.   For small $z$
the function  $K_\phi(z)$  takes the the asymptotic value  $K_\phi  \to  1$.
The  quantity $\zeta_\phi(E)$
can be understood qualitatively as 
 the number of  Hubble times 
that ``count'' in producing  the  observed  inclusive flux.
In the case of a  ``non--interactive''  particle like the neutrino
$\zeta_\phi(E)$ is  clearly of order unity  
because the age of our universe is in fact of order of one Hubble time,
In  case of strong source evolution, with a  much larger  activity
in the past, the factor $\zeta_\phi$ can be as large as 2--3.
For  ultra--high energy  protons 
 the value of $\zeta_\phi(E)$ reflects  (in units  of the Hubble
time  $H_0^{-1}$)  the   characteristic time
for energy loss  of the  particles
and   $\zeta_\phi \ll 1$,  varying also  rapidly with energy.

The  appearance  in equation (\ref{eq:ndens})   of the  function  $F_{\phi}(z,E)$  (that is 
dimming function for  a  source calculated  assuming 
linear propagation for  the emitted particles) 
may appear surprising.
It can   therefore  be instructive to give  a  second derivation of the 
equation (\ref{eq:ndens})   assuming  that the particles
propagate  along straight lines. In this case, if the particles
move with a  velocity close to the speed of light, and  the 
distance between sources is small  with respect to the Hubble  length $c/H_0$  the
condition of homogeneity  of the  emitted particle  density  is  clearly  satisfied.
For particles  propagating at the speed  of light along  straight lines
the redshift $z$ of  the emission of a particle
is not only in  a  one--to--one  relation with cosmic  time, 
but also with the source  distance.
The inclusive flux  can therefore 
be computed summing   the contributions
of  the  sources  contained in   different    layers
  of  comoving volume.
The  inclusive flux  is  then 
obtained  with the double  integration:
\begin{equation}
\phi_{\rm incl}(E) = \frac{1}{4 \, \pi} ~
\int_0^\infty dz~\frac{dV}{dz} ~ \int dL~\frac{dn_s}{dL}(L,z)
~\phi(E,L,z) 
\end{equation}
where    $dV/dz$  is the comoving  volume  contained
in the layer $[z,z+dz]$
\begin{equation}
\frac{dV(z)}{dz} = 4\, \pi \; R_0^3 ~ \frac{r^2(z)}{ {\cal H}(z)} ~,
\label{eq:dvol}
\end{equation}
$dn_s/dL(L,z)$ is the number  density of  sources   with luminosity
$L$  at the epoch $z$, and $\phi(E,L,z)$ is the flux
from a  source of luminosity $L$ placed at distance $z$.
For  a universal  emission shape $q(E)$ (normalized to
have the median luminosity $L_0$)  one has:
\begin{equation}
\phi(E,L,z) = \frac{q(E)} {4 \pi \; R_0^2} 
~\frac{L}{L_0} ~F_\phi(z,E)
\end{equation}
with $F_{\phi}(z,E)$ the dimming function   defined in (\ref{eq:f_gen}).
It is  then  simple to see that one reobtains
expression (\ref{eq:ndens}), confirming the  consistency of the treatment.

It is also possible
to  use an analogous notation  for the inclusive integral  flux
introducing the functions $K_{\Phi}(z,E)$ and $\zeta_\Phi(E)$:
\begin{equation}
\Phi_{\rm incl} (E_{\rm min}) 
= \frac{R_0}{4 \, \pi} ~
Q_{\rm incl}(E_{\rm min},0) 
~ \zeta_\Phi(E_{\rm min})
= \frac{R_0}{4 \, \pi} ~
\frac{{\cal L}(0)  \; \eta}{E_{\rm min}} 
~ \zeta_\Phi(E_{\rm min})
\label{eq:Phi_incl}
\end{equation}
with 
\begin{equation}
\zeta_\Phi(E_{\rm min}) = \int_0^\infty dz 
~ K_{\Phi}(z,E_{\rm min}) 
= \int_0^\infty dz ~
\frac{r^2(z)}{ {\cal H}(z)} ~ F_{\Phi}(z,E_{\rm min})
 ~\frac{ {\cal L}(z)} {{\cal L}(0)}  
\label{eq:xib}
\end{equation}
where  the dimming function $F_{\Phi}(z, E_{\rm min})$ for  the  integral flux  was
defined in equation (\ref{eq:f_int}).

\begin{figure} [hbt]
\centerline{\psfig{figure=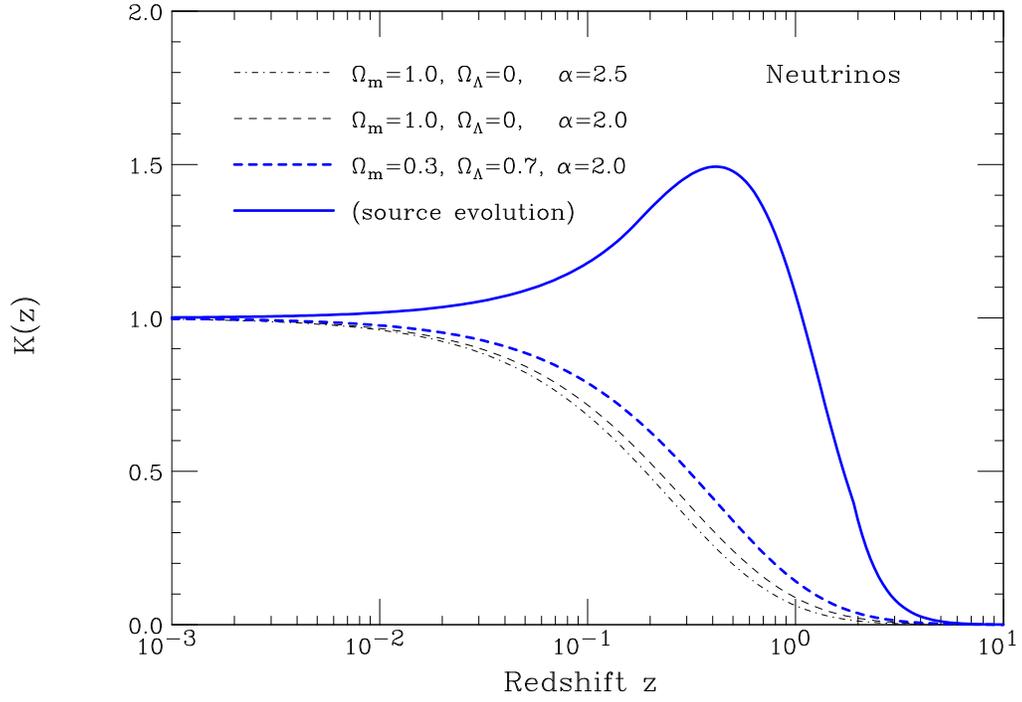,angle=90,width=13.5cm}}
\caption {\footnotesize 
Plot of the 
redshift response (``Kepler--Olbers'') 
 function $K_\phi(z) = K_\Phi(z)$  for  neutrinos
assuming a power law  emission of slope $\alpha$.
In three of the calculations the  neutrino emission
is  considered constant in time. In  a fourth calculation
the neutrino emission   changes in time
with the  redshift dependence shown in  fig.~\protect\ref{fig:lum_evol}.
\label{fig:nu2}  }
\end{figure}

Some examples of the redshift response (or Kepler--Olbers)  function
for    neutrinos  emitted with  a
power  law  spectrum  of slope $\alpha$
($q(E) \propto  E^{-\alpha}$)   are shown  in 
figure~\ref{fig:nu2}. In this  particular case 
(no  energy loss except for the redshift effects, and
a featureless injection spectrum)  the
redshift response is   independent from energy and  one has:
\begin{equation}
K_\phi(z) = K_{\Phi}(z) = \frac{(1+z)^{-\alpha}}{{\cal H}(z)} ~ 
 ~\frac{ {\cal L}(z)} {{\cal L}(0)} 
\end{equation}
As  a consequence also the ``shape  factors''
$\zeta_\phi = \zeta_\Phi$ are equal to each other and independent from energy.
This  means  that an injected  power law spectrum of neutrinos
is not deformed  by propagation effects.
As numerical examples,  
for a  matter dominated  cosmology
($\Omega_{\rm m} = 1$, 
 $\Omega_{\Lambda} = 0$)   and  no  source  evolution 
one  has:
\begin{equation}
\zeta_\phi (\alpha) =
\zeta_\Phi (\alpha) = \frac{2}{2 + \alpha}
\end{equation}
 Softer  spectra  (larger slope $\alpha$)   have  a smaller
value of $\zeta_\phi$.
For the  concordance   cosmology
($\Omega_{\rm m} = 0.3$, 
 $\Omega_{\Lambda} = 0.7$)   the 
shape  factors   are larger  by approximately  30\%,
for  example for $\alpha = 2$   and  no evolution
$\zeta_\phi = 0.5326$.
The inclusion of     source  evolution can  on the other  hand
increase  the  value of the shape  factor by a  
significant factor. 
For example  using  the same
redshift  dependence  of the 
AGN X--ray  luminosity of  shown  in figure~\ref{fig:lum_evol}  the 
shape factor for neutrinos  increases by  a factor $\simeq  4$
(for $\alpha = 2$  one  finds $\zeta_\Phi \simeq 2.2$).

\begin{figure} [hbt]
\centerline{\psfig{figure=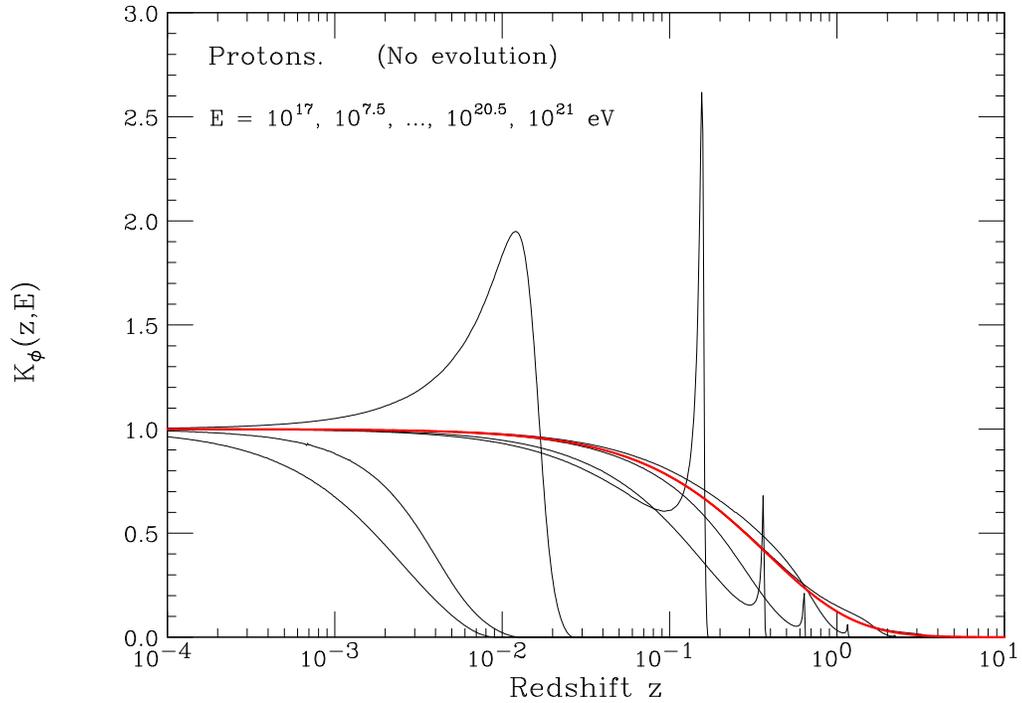,angle=90,width=13.5cm}}
\caption {\footnotesize 
Plot of the 
redshift response (``Kepler--Olbers'') 
 function $K_\phi(z,E)$  for  protons 
assuming a power law  emission of slope $\alpha=2.3$.
The different curves are calculated (in the absence of
source  evolution)  for  different values
of the observed  proton energy.
The thick (red)  curve is the  curve  calculated
for neutrinos (considering  only   redshift energy losses).
\label{fig:k_prot}  }
\end{figure}

\vspace{0.3 cm}
The  redshift response for  protons has  a much more  complex
and interesting structure.  Figure~\ref{fig:k_prot}
shows the redshift response function $K_\phi(z,E)$ for  values of  the 
proton energy.  One  can see that for  sufficiently low energy $E \lesssim 10^{17}$~eV,
the function   takes the same  form as for neutrinos,  since   the role
of $p\gamma$ interactions  becomes  negligible.  With increasing energy 
the range  of redshift that  contribute to the flux  becomes  progressively smaller.
Note also how the   energy loss  effects  enhace  the role of  a narrow range  of  redshifts
where the high energy tail of the emission  rapidly lose energy.

\begin{figure} [hbt]
\centerline{\psfig{figure=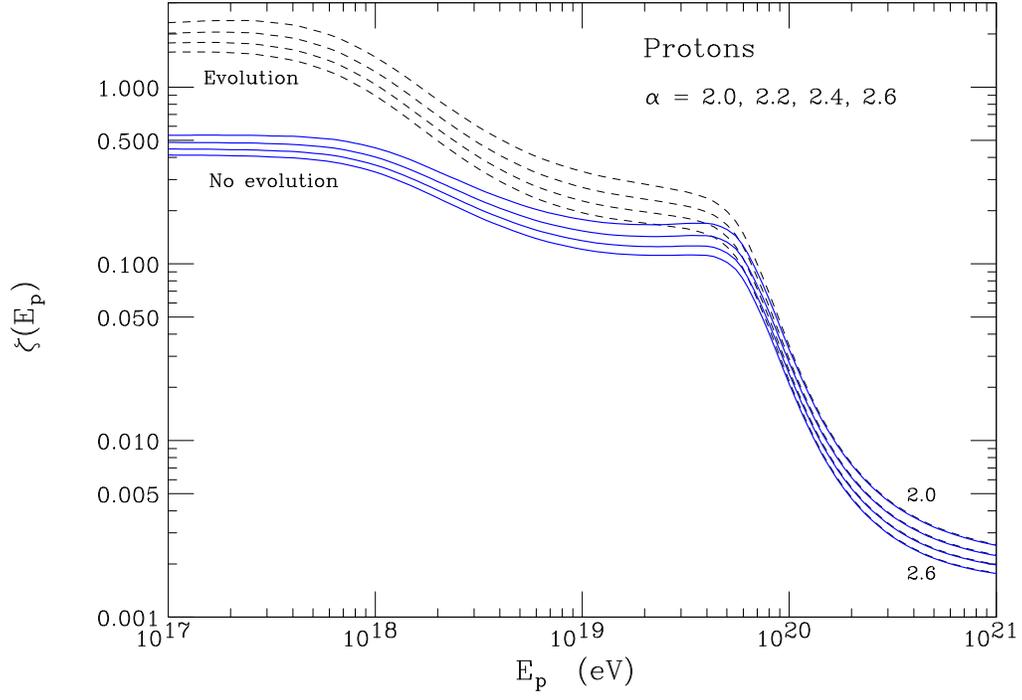,angle=90,width=13.5cm}}
\caption {\footnotesize 
Shape  factor  $\zeta_\phi(E)$  for  protons, calculated  for
a  power  injection with  slope
$\alpha = 2.0$, 2.2, 2.4 and~2.6.
 The solid lines are calculated  for a constant value
of the  source  power density, while the dashed  lines assume
for ${\cal L}_p(z)$ the   redshift  dependence  shown 
in  fig.~\protect\ref{fig:lum_evol}.
\label{fig:csi}  }
\end{figure}

Figure~\ref{fig:csi}  shows the energy dependence of the factor $\zeta_\phi(E)$. 
For  energies much smaller that  $E \simeq  10^{18}$~eV the   factor $\zeta_\phi(E)$  reaches an asymptotic
value   that is  in fact equal to the one for neutrino,
while at larger  energies  the effect of
$\zeta_\phi(E)$ is to  distort  the power law  injection spectrum.
The factor  $\zeta_\phi(E)$ plays therefore  the role of a ``shape factor''  
that  marks  the  ``imprints''  on  energy loss
on the observable energy spectrum.
Berezinsky and his collaborators    \cite{Berezinsky:2005cq}  have discussed how this 
spectral  distortion  distortion
could be  the origin  of the ``ankle''  structure  observed in the CR flux.

\begin{figure} [hbt]
\centerline{\psfig{figure=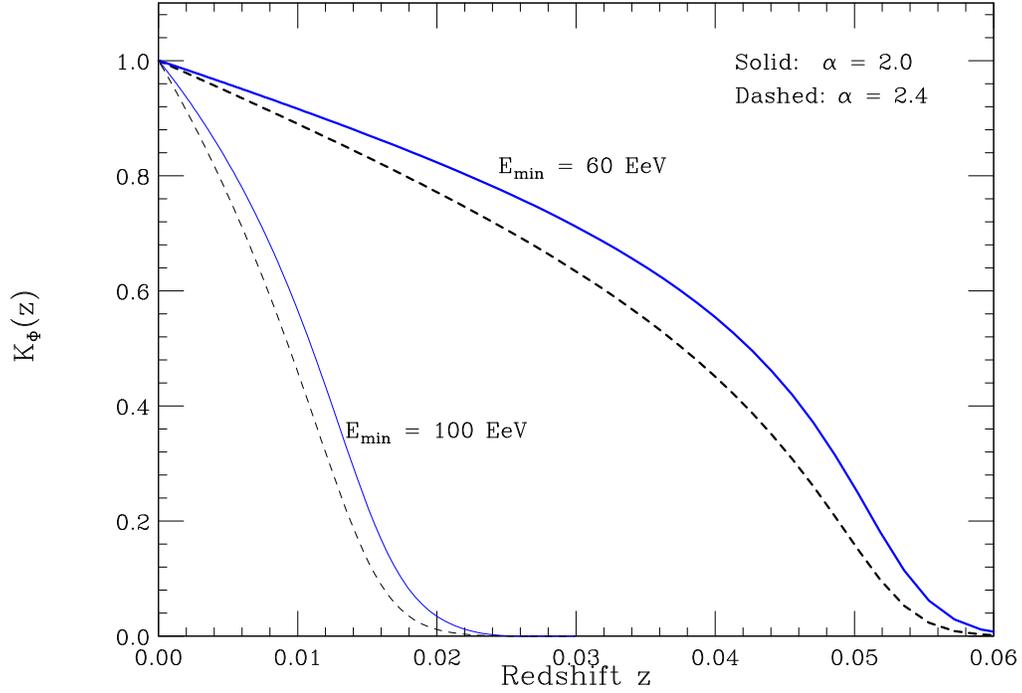,angle=90,width=13.5cm}}
\caption {\footnotesize 
Redshift dependence of the Kepler--Olbers  functions
$K_\Phi(z, E_{\rm min})$    for   
$E_{\rm min} = 6 \times 10^{19}$~eV  (thick  curves)
and
$E_{\rm min} = 10^{20}$~eV  (thin  curves).
The  solid (dashed) curves  assume  a power law
 injection spectrum with a slope $\alpha = 2.0$ ($\alpha = 2.4$).
Integrating over  $z$  one obtain  the values
$\zeta_\Phi  =  0.037$ (0.034) for $E_{\rm min} = 6 \times 10^{19}$~eV
and
$\zeta_\Phi  =  0.0107$ (0.0093) for $E_{\rm min} =10^{20}$~eV.
%(The smaller  number refer to the softer spectrum). 
\label{fig:kepl_60}  }
\end{figure}

Figure~\ref{fig:kepl_60}  shows
the  Kepler--Olbers function  $K_{\Phi}(z,E_{\rm min})$  
for the integral  flux for two values of $E_{\rm min}$
($E_{\rm min} = 6 \times 10^{19}$~eV   and
$10^{20}$~eV);  the proton injection spectrum 
is  a power law with slope $\alpha = 2.0$ and~2.4.
 At these very high  energies  
only a small     volume of the universe  is 
observable, and the size of this  observable volume
changes very rapidly with the threshold energy.
Integrating over  $z$  one  obtains  (for $\alpha = 2.0$)
 $\zeta_\Phi  =  0.037$ 
for  $E_{\rm min} = 6 \times 10^{19}$~eV and
$\zeta_\Phi  =  0.0107$  for $E_{\rm min} =10^{20}$~eV.
For the softer spectra 
($\alpha = 2.4$) 
$\zeta_\Phi$  shrinks   by approximately 10\%.

%\subsection{Particle horizon}
\vspace{0.4 cm}
The calculation of the  inclusive flux 
as an integral over  redshift:
\begin{equation}
\phi(E) = \frac{R_0}{4 \, \pi} ~q_{\rm incl}(E)~ \int_0^\infty dz~ K_\phi(z,E)
 = \frac{R_0}{4 \, \pi} ~q_{\rm incl}(E)~\zeta_\phi(E)
\end{equation}
suggests  the most useful  definition  of an energy
dependent  redshift horizon 
using the  implicit equation:
\begin{equation}
\int_0^{z^{\rm  hor}_{f}} dz~K_\phi(z,E)
= f~\zeta_\phi(E)
\label{eq:zhor}
\end{equation}
For example $z^{\rm hor}_{0.9}(E)$  is  the redshift 
that contains  the emission of  90\% of the protons  that are
observed with  energy $E$.   
Note that this  definition    of the horizon depends  on the 
assumed  shape  of the emission spectrum.

\begin{figure} [hbt]
\centerline{\psfig{figure=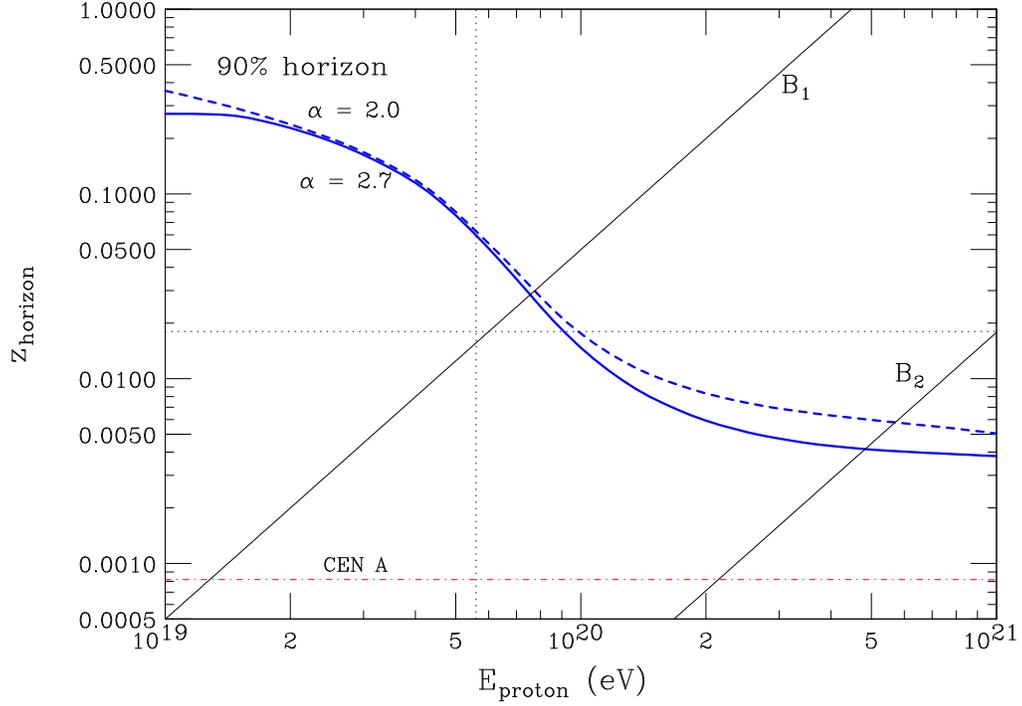,angle=90,width=13.5cm}}
\caption {\footnotesize 
The thick  curves show the redshift horizon as  a function
of the proton energy.  The horizon is defined
as the  value of $z$  within which 90\%  of the  observed
flux is   emitted.
The solid (dashed) line  is  calculated  for a power law emission
with slope $\alpha = 2.7$ (2.0).  The  dotted  curves
indicate the selection criteria used by the Auger 
collaboration in  their correlation analysis \protect\cite{Cronin:2007zz}.
The  redshift that corresponds to the Cen~A position is also indicated.
The  curves  labeld $B_1$ and $B_2$ correspond
to an  expected  deviation of 3 degrees with the estimates
of the extragalactic  magnetic  field of equation
(\protect\ref{eq:estim1})
and
(\protect\ref{eq:estim2}).
\label{fig:hor1}   }
\end{figure}

Figure~\ref{fig:hor1}  shows the   90\% redshift  horizon as a function
of the proton energy, assuming injection with  a power  law 
of slope $\alpha$.  In the same  figure we show also the   energy dependence 
of the imaging radius (for a 3$^\circ$ resolution)   calculated 
for the  two estimates of the   parameters $B \sqrt{\lambda_B}$  
of equations (\ref{eq:estim1}) and
(\ref{eq:estim2}).   Proton  astronomy is  going to be limited to  a region
$z < z_{\rm hor}(E)$ and $z < z_{\rm magnetic} (E) \simeq R_{\rm  imaging}(E)/R_0$.

\section{The  Intensity and Multiplicity  distributions}
If the  universe is filled  with a   homogeneous 
distribution of  sources,    an  observer  will  see 
an  ensemble of  objects  with a  
broad  distribution  of   flux intensity, 
with the faintest (brighest) objects corresponding to  the 
most  distant (closest)  sources.  

In the following we will consider the
the  integral  flux  above a   fixed  threshold
energy $E_{\rm min}$. In the following expressions
the  dependence on this minimium energy will be left implicit.
If all sources  have the same spectral shape,
the  distribution  $dN_s/d\Phi$  that  gives   the number
of  sources   observable  with  flux  in the interval
$[\Phi, \Phi + d\Phi]$   can be written in the   form:
\begin{equation}
\frac{dN_s}{d\Phi} = 
\frac{N_{\rm Hubble}}{\Phi_0}~
~ G_\Phi \left ( \frac{\Phi}{\Phi_0} \right )
\label{eq:dndf}
\end{equation}
where  the  adimensional constant $N_{\rm Hubble}$:
\begin{equation}
N_{\rm Hubble} = 4 \, \pi \; R_0^3 ~
\frac{ {\cal L}(0)} {L_0} 
\label{eq:nh0}
\end{equation}
is proportional to the number of sources in  one Hubble  volume,
and the   flux  $\Phi_0$:
\begin{equation}
\Phi_0 = \frac{Q_0}{4 \pi \, R_0^2} =
 \frac{L_0~\eta}{4 \, \pi \; R_0^2 \; E_{\rm min} }
\label{eq:ff0}
\end{equation}
is the  flux   from a   ``typical source'' 
(chosen here  as the source with the median luminosity $L_0$)
    placed
at the Hubble distance  and   calculated   neglecting 
all  cosmological and energy loss effects.
The  general  expression for $G_\Phi (x)$ is:
\begin{equation}
G_\Phi (x) = 
 \int_0^\infty dz  ~\frac{r^2(z)}{{\cal H}(z)} ~
~\int_0^\infty dy~ f_L (y,z) 
~ \delta \left [x - y  ~F_{\Phi}(z)  \right ]~.
\label{eq:gPhi}
\end{equation}
In this  expression  $r(z)$ and ${\cal H}(z)$  depend on the
cosmological parameters, $F_{\Phi}(z)$   is the ``dimming function''
(defined in  equation (\ref{eq:Phi_gen}))  and
$f_L(y,z)$ is the (scaled)  luminosity functions (see equation
(\ref{eq:flum})).
The  function $G_\Phi (x)$   encodes 
the shape  of the emission spectrum,
the properties of  particle  propagation and the 
luminosity function and  cosmic  evolution of the sources.

As  a simple check of this  expression 
one can compute the inclusive  particle  flux 
is  obtained  integrating over all  fluxes $\Phi$ the distribution:
\begin{eqnarray}
\Phi_{\rm incl}  & = & 
\frac{1}{4 \, \pi}
\int_0^\infty d\Phi~ \Phi~\frac{dN_s}{d\Phi} 
= 
N_{\rm Hubble} 
~\Phi_0
~ \int_0^\infty dx~x~G_\Phi (x) 
%\nonumber  \\
%& ~ & \nonumber \\
%& = &  \frac{R_0}{4 \, \pi} ~ \frac{ {\cal L}(0) \; \eta}{E_{\rm min}}
%~ \int_0^\infty dx~x~G_\Phi (x)
\end{eqnarray}
This  expression is  identical  to the result  (\ref{eq:Phi_incl}) 
because the integral over $x$  $x\; G_\Phi(x)$    is equal to $\zeta_\Phi$.
In fact, 
using the definition of  $G_\Phi(x)$ of equation  (\ref{eq:gPhi}) one 
 finds:
\begin{eqnarray}
\int_0^\infty dx~x~G_\Phi (x) &  = &
\int_0^\infty dz~\frac{r^2(z)}{ {\cal H}(z)} 
~F_{\Phi}(z)~\left ( \int_0^\infty dy~y~f_L(y,z) \right )
\nonumber \\
& ~ & \nonumber \\
& = &  
\int_0^\infty dz~\frac{r^2(z)}{ {\cal H}(z)} 
~F_{\Phi}(z)
~\frac{{\cal L}(z)}{{\cal L}(0)}
 = \zeta_\Phi
\label{eq:gf_int}
\end{eqnarray}

To be in closer contact with
the observations, we will    consider an explicit
model of  detector,
described  by  the effective area $A(\Omega)$ 
and  a  finite  observation time $t$ and
 study the  number  of sources $N_k$ 
detected with exactly  $k$  (an  integer  quantity)  events.
 The multiplicity distribution $N_k$   can be  obtained
starting from the  function   $dN_s/dm$,  where 
$m$  (a  real, continuosly  varying quantity) 
is the expected  signal  from a source. The
quantity $dN_s/dm$  
gives  the number of sources  with   expected signal
in the interval $[m,m+dm]$.
The  probability to  detect
$k$ events from a source
with expected signal  $m = \langle k\rangle$ 
can be obtained   using Poissonian  statistics.
The   expected  number of events  with
multiplicity $k$ is therefore:
\begin{equation}
N_k = \int_0^\infty dm~ \left ( 
e^{-m} ~\frac{m^k}{k !} \right )~ \frac{dN_s}{dm}  ~,
\label{eq:singlets}
\end{equation}
the number of  multiplets  with multiplicity $k\ge 2$  is:
\begin{equation}
N_{\rm multi} \equiv 
N_{\ge 2} = \int_0^\infty dm~[1 -  e^{-m} \; (1+m)]\; \frac{dN_s}{dm} 
\label{eq:multi0}
\end{equation}
and the total  number of  {\em detected} sources  is:
\begin{equation}
N_{\rm sources} \equiv  N_{\ge 1} = \int_0^\infty dm~[1 -  e^{-m}]\;
 \frac{dN_s}{dm} 
\label{eq:sources}
\end{equation}
For the calculation of  the function $dN_s/dm$
we can  observe that the expected  signal $m$  from a source 
of luminosity $L$,  redshift $z$ and   celestial 
coordinates $\Omega$ is:
\begin{equation}
m(L,z, \Omega)   = \Phi(L,z)  \; A(\Omega) \, t  = 
~\Phi_0 ~\frac{L}{L_0} 
~F_{\Phi}(z)  \; A(\Omega) \, t  = 
\frac{L \; \eta}{4 \, \pi \; R_0^2  \; E_{\rm min}}\; F_{\Phi}(z) \; A(\Omega) \, t 
\end{equation}
%where $\Phi$ is the  source  flux, 
% $A(\Omega)$ is the detector
%average effective area for the source  direction $\Omega$ and 
%$t$ is the  measurement live time. 
The   complication here is that in  general 
 the detector  exposure     depends  on the celestial coordinates.
A simple   modeling   of the average   effective  area
of the Auger  detector and of  neutrino  telescopes as
a function of celestial declination is discussed in appendix~\ref{sec:app_eas}.
We will  use the notation:
\begin{equation}
A(\Omega) = \langle A \rangle \; a(\Omega)  = \frac{A_\Omega}{4 \, \pi} ~a(\Omega)
\label{eq:aomega}
\end{equation}
where $\langle A \rangle$ is the detector  average    effective  area,
$A_\Omega$ is  the angle  integrated  effective area
(in units cm$^2$~sr)   and $a(\Omega)$ is a dimensionless function
that satisfies the normalization condition:
\begin{equation}
\int \frac{d\Omega}{4 \, \pi} ~ a(\Omega) = 1
\end{equation}
The  distribution  $dN_s/dm$   can then be  written as:  
\begin{equation}
\frac{dN_s}{dm} = 
\frac{N_{\rm Hubble}}{m_0}~
~ G_m \left ( \frac{m}{m_0} \right )
\label{eq:dndm}
\end{equation}
with  $N_{\rm Hubble}$  defined in  equation (\ref{eq:nh0})  and:
\begin{equation}
m_0 = 
\Phi_0 \; \langle A \rangle \; t = 
\frac{L_0~\eta}{4 \, \pi \; R_0^2 \; E_{\rm min} } ~\langle A \rangle \;t
\label{eq:mm0}
\end{equation}
The ``scaling structure'' of
equations  (\ref{eq:dndf}) and (\ref{eq:dndm}) can be 
easily deduced  with  simple dimensional analysis
observing also  that   the number of sources  must be proportional
to the quantity  ${\cal L}(0)/L_0$  that is itself  proportional  
to the source  density,
and that the signal from any
source is  linear in the  observation time $t$, 
the detector size (proportional  to $\langle A \rangle$), and 
the source luminosity $L$.
 
The  function $G_m (x)$ can be  calculated  as the convolution:
\begin{equation}
G_m(x) = 
 \int \frac{d\Omega}{4 \, \pi} ~\frac{1}{a(\Omega)}
~ G_\Phi \left (  \frac{x}{a(\Omega)} \right )
\end{equation}
Another way to express the same  result is:
\begin{equation}
G_m(x) = 
 \int \frac{d\Omega}{4 \, \pi} 
~ \int_0^\infty dz  ~\frac{r^2(z)}{{\cal H}(z)} ~
~\int_0^\infty dy~ f_L (y,z) 
~ \delta \left [x - y \; a(\Omega) ~F_{\Phi}(z)  \right ]
\label{eq:ggm_def}
\end{equation}
The function $G_m(x)$  satisfies the  ``sum rule'':
\begin{equation}
\int_0^\infty dx ~x~G_m(x) = 
\left (\int  \frac{d\Omega}{4 \, \pi} ~a(\Omega)  \right )
~\left (\int_0^\infty dx ~x~G_\Phi(x) \right )
 =  \int_0^\infty dx ~x~G_\Phi(x) = \zeta_\Phi 
\end{equation}
As  a consistency check one  can 
compute the total  number of detected  events 
as:
\begin{eqnarray}
N_{\rm events} & =  & 
\int_0^\infty dm~m ~ \frac{dN_s}{dm}
 =  
N_{\rm Hubble} ~m_0 ~ \int_0^\infty dx~x~G_m(x) 
\nonumber \\
& ~ & \nonumber \\
& = & 
\left ( \frac{R_0}{4 \,  \pi }~\frac{ {\cal L}(0) \; \eta}{E_{\rm min}}
 ~\zeta_\Phi \right )
~A_\Omega \; t
= \Phi_{\rm incl} ~  A_\Omega \; t 
\label{eq:nev}
\end{eqnarray}
The  result  is  simply  the inclusive flux
(already calculated    in   equation 
 (\ref{eq:Phi_incl}))   times  the total  detector
exposure  $A_\Omega \; t$, as it must be.
Note that  the  total number of detected  events 
does  not depend  on the luminosity function of the sources,
but only on the  (space  averaged)  emissivity, and also 
that the result is  independent from the 
geometry or  ``field of view'' of the detector.
%If we have two  identical  detectors, one at the South  Pole, and a second
%at the equator,  their  inclusive    event rate is   identical,
%even if the South Pole detector   observes  
%more ``deeply''  a smaller portion of the celestial sphere.

The shape of the functions  $G_\Phi(x)$ 
or $G_m(x)$ in general is complicated,
since it encodes 
 the  properties of particle propagation, the source
luminosity function  and, in the case of  $G_m(x)$,   also the 
detector  geometry,  however the limit of  large 
$x$  (that is  the largest  fluxes or  signals)  the   functions have 
  simple asymptotic  behaviours:
\begin{equation}
G_{\Phi}(x) \longrightarrow \frac{x^{-5/2}}{2} ~ \langle y^{3/2} \rangle
\end{equation}
and 
\begin{equation}
G_{m}(x) \longrightarrow \frac{x^{-5/2}}{2} ~ \langle y^{3/2} \rangle
 ~ \langle a^{3/2} \rangle
\label{eq:g_asy}
 \end{equation}
 where the quantity
\begin{equation}
\langle y^{3/2} \rangle =
\int_0^\infty dy ~y^{3/2} ~f_L (y, 0)
\label{eq:y32}
\end{equation}
depends on  the luminosity function, and 
\begin{equation}
\langle a^{3/2} \rangle =
\int \frac{d\Omega}{4 \, \pi} ~a^{3/2}(\Omega) 
\end{equation}
depends on the detector location and  geometry.
These physically intuitive   results can be 
easily  obtained  computing the  euclidean/static  limits
of  equations 
(\ref{eq:gPhi}) 
and (\ref{eq:ggm_def})  with  three steps:
(i)  use  the  asymptotic  (small $z$  expression) 
$r^2(z)/{\cal H}(z) \to z^2$; (ii)  approximate
the function  $F_{\Phi}(z) \to z^{-2}$; and
(iii) neglect the redshift dependence of the
scaled  luminosity function $f_L(y,z) \to f_L(y)$.
The  integral over redshift  can then  be performed
using  the delta function, and:
\begin{equation}
\int dz ~ z^2 ~\delta \left [ x - \frac{a}{z^2} \right ]
 = \left [ z^2 ~ \left |\frac{2 \,a }{z^3} \right |^{-1}
 \right ]_{z = \sqrt{a/x}} 
= \frac{a^{3/2}}{2} ~x^{-5/2} 
\end{equation}
The $m^{-5/2}$  ($\Phi^{-5/2}$)  behaviour of the distribution 
of expected signal  (flux) remains  valid
only for sufficiently  large signals  (fluxes),
then  the distribution stops  growing so
quickly for    decreasing $m$, and the 
corrected  distribution is  integrable   down to  $m\to 0$.

The  number of sources  with and 
expected  signal  larger  than   the  minimum value
$\bar{m}$  can be  calculated integrating the distribution $dN_s/dm$:
in the  interval  $ m \ge \bar{m}$. If $\bar{m}$ is
sufficiently large one  can use the asymptotic
form (\ref{eq:g_asy})  with the result:
\begin{equation}
N_s (\ge \bar{m}) =
\int_{\bar{m}}^\infty dm~\frac{dN_s}{dm} 
\simeq 
 ~\frac{4 \, \pi}{3}
 \frac{{\cal L}(0)} {L_0}
~R_0^3~
\left ( \frac{m_0}{\bar{m}} \right )^{3/2}
%\left (\frac{L_0 \; \eta ~\langle A \rangle \, t}{4 \, \pi \; E_{\rm min}}
%~\frac{1}{\overline{m}} \right )^{3/2}
~\left  \langle a^{3/2} \right \rangle
~\left  \langle y^{3/2} \right \rangle
\label{eq:nmm}
\end{equation}
This  result has  a very transparent  meaning
if one  considers the limit  of  isotropic  acceptance  and  
identical  source luminosity;  then   the factors 
$\langle a^{3/2} \rangle$ and
$\langle y^{3/2} \rangle$  become  unity,
the quantity ${\cal L}/L_0$   is  simply the   source  density,
and the  result  takes the form
\begin{equation}
N_s (\ge m)  \to 
n_s ~\frac{4 \, \pi}{3} ~ R_m^3 
\end{equation}
where 
\begin{equation}
R_m = R_0~
\left (\frac{m_0}{\bar{m}} \right )^{1/2}
= \left ( \frac{L_0 \; \eta ~\langle A \rangle \, t}{4 \, \pi \; E_{\rm min}}
~\frac{1}{m} \right )^{1/2}
\end{equation}
is the radius  (in  static, euclidean space)  of the sphere
at wich the signal  has  value  $m$.  Note that  the radius  $R_m$ 
is  independent from the Hubble distance $R_0$, as  it must be,
since it is calculated  neglecting
cosmological effects.

Similarly, the expected number of  sources
that yield  a ``multiplet''  of multiplicity $k \ge 2$ can 
be  calculated using
the asymptotic  form of   $G_m$ with the result:
\begin{equation}
N_{\rm  multiplets}   =
\int_0^\infty dm ~[1 - e^{-m}\, (1+m)]  ~ \frac{dN_s}{dm}
\simeq    \sqrt{\pi} ~ N_s (m \ge 1)
\label{eq:nmult0}
\end{equation}
where   we have used the  result:
\begin{equation}
\int_0^\infty  dm~[1 - e^{-m} \, (1+m)]~\frac{m^{-5/2}}{2} 
= \frac{\sqrt{\pi}}{3} ~
\end{equation}
The expected  number of multiplets can  be well
approximated  as
 the number of  sources  with an expected signal
larger  than  unity, multiplied  by a factor
$\sqrt{\pi}$. 
The limit of  validity of the  approximation (\ref{eq:nmult0})  will
be  discussed    later.

\begin{figure} [hbt]
\centerline{\psfig{figure=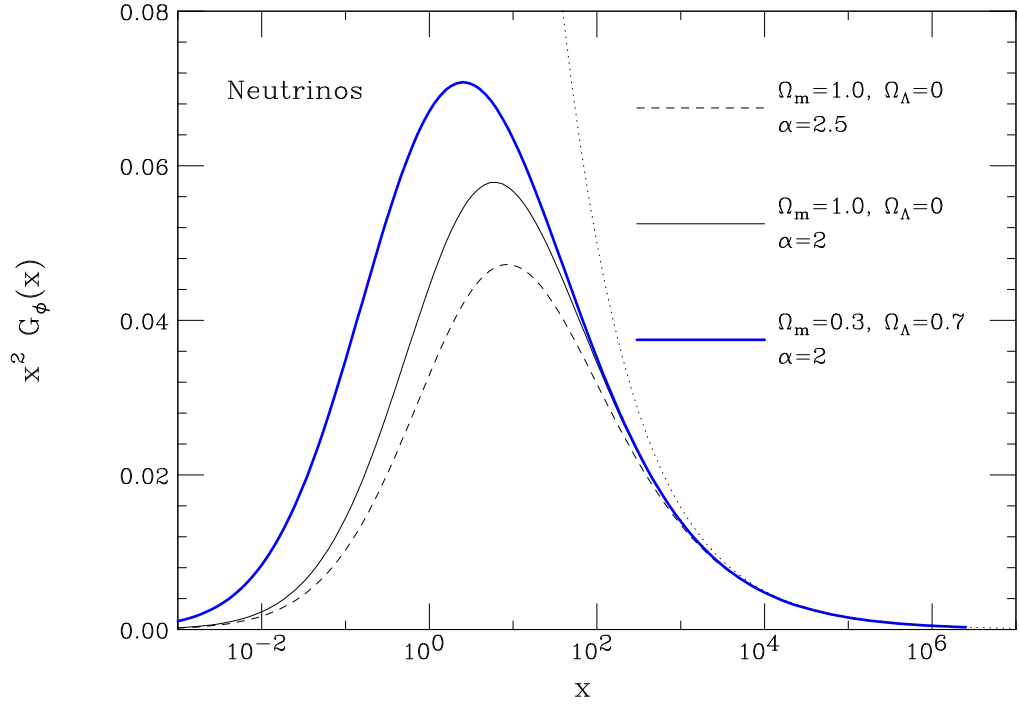,angle=90,width=13.5cm}}
\caption {\footnotesize 
Plot of the function  $G_\phi(x)  = G_\Phi(x)$  
for neutrinos  observations.  The  neutrino  emission has a power
law  form with slope $\alpha$, and all neutrino sources
have identical luminosity.
The different curves  are calculated   for different
cosmological parameters, and for two different
values of the slope $\alpha$.
The dotted  curve show the function $x^{-5/2}/2$.
\label{fig:nu3}  }
\end{figure}

Some examples of   the function
$G_\Phi(x)$  calculated for neutrino  sources
assuming a power law  emission, and  all identical  sources
are shown in fig.~\ref{fig:nu3}.   The  function
$G_\Phi(x)$ is calculated  for two different 
set of values of the  cosmological parameters 
(a matter  dominated universe 
and the concordance cosmology)
and two  different  value  of  the slope $\alpha$.
One can  see that for $x$ large the function  takes the
``universal shape'' $G_\Phi(x)  \to x^{-5/2}/2$, while for sufficiently
small $x$  the curve  deviates from the ``euclidean form''.
The  small $x$  (or faint source part)   of the function
is  not universal, but depend on the cosmological parameters
and the shape of the emission spectrum.

\begin{figure} [hbt]
\centerline{\psfig{figure=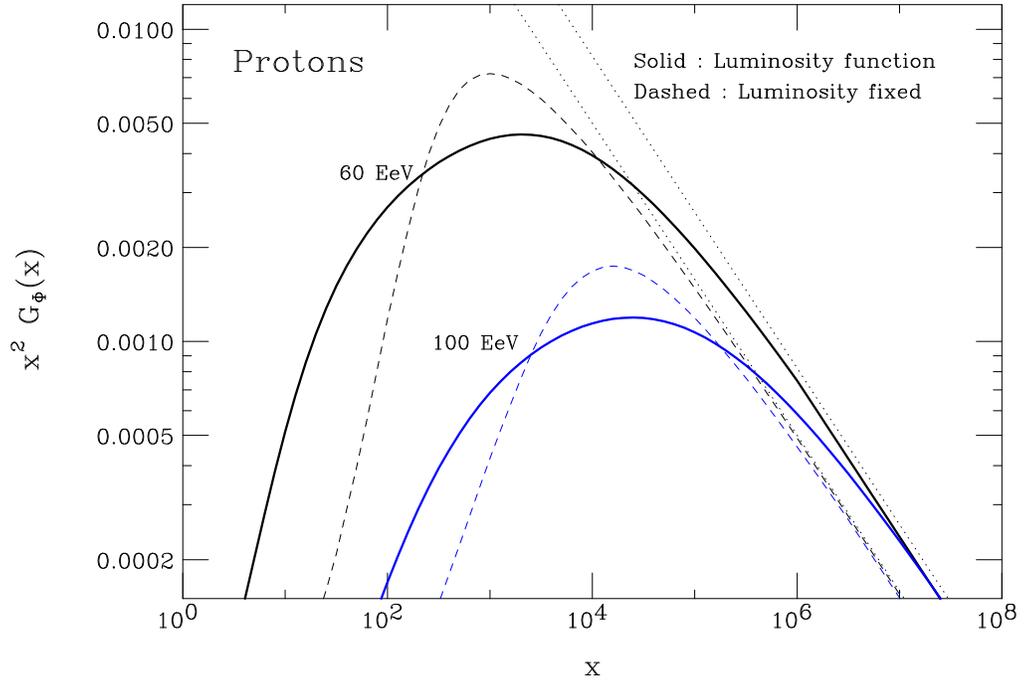,angle=90,width=13.5cm}}
\caption {\footnotesize 
Plot of the function $G_\Phi(x)$  
(in the form  $x^2 \;G_\Phi(x)$  versus $x$)  for the
proton integral flux.
The calculation  is performed  for a power
law energy spectrum with $\alpha = 2.4$ and
two energy threshold 
$E_{\rm min} = 6 \times 10^{19}$~eV
and 
$E_{\rm min} = 10^{20}$~eV.
The dashed  lines are calculated  assuming that  all sources
have the same  luminosity,  while the solid line
assume the luminosity function of  equation
(\protect\ref{eq:lum_shape})  shown also in  fig.~\ref{fig:lum_shape}.
The dotted  curves  show the functions $x^{-5/2}/2$ 
and $x^{-5/2} \, \langle y^{3/2} \rangle /2$ 
\label{fig:gg_p}  }
\end{figure}

Figure~\ref{fig:gg_p}  shows  some examples   of the 
$G_\Phi(x)$   function calculated for protons,  
integrating the observed flux   above two
different   threshold energies
$E_{\rm min} = 6 \times 10^{19}$~eV and
$E_{\rm min} = 10^{20}$~eV. 
The  calculation  is performed  for an emission spectrum
$E^{-2.4}$  with two assumptions about the sources,
either all  identical  sources, or the broad  
luminosity  distribution of equation  (\ref{eq:lum_shape}).
Again, at large  $x$   the function takes its  universal form
$x^{-5/2}/2$   for a fixed source luminosity
or  
$x^{-5/2}  \langle y^{3/2}\rangle/2$ 
for  a luminosity distribution.
However for protons   this euclidean behaviour  stops
at a  larger  $x$, because  of the effect of energy loss.
In fact for   the higher energy threshold,  the  euclidean
behaviour is  valid in  a smaller  range of $x$.

\begin{figure} [hbt]
\centerline{\psfig{figure=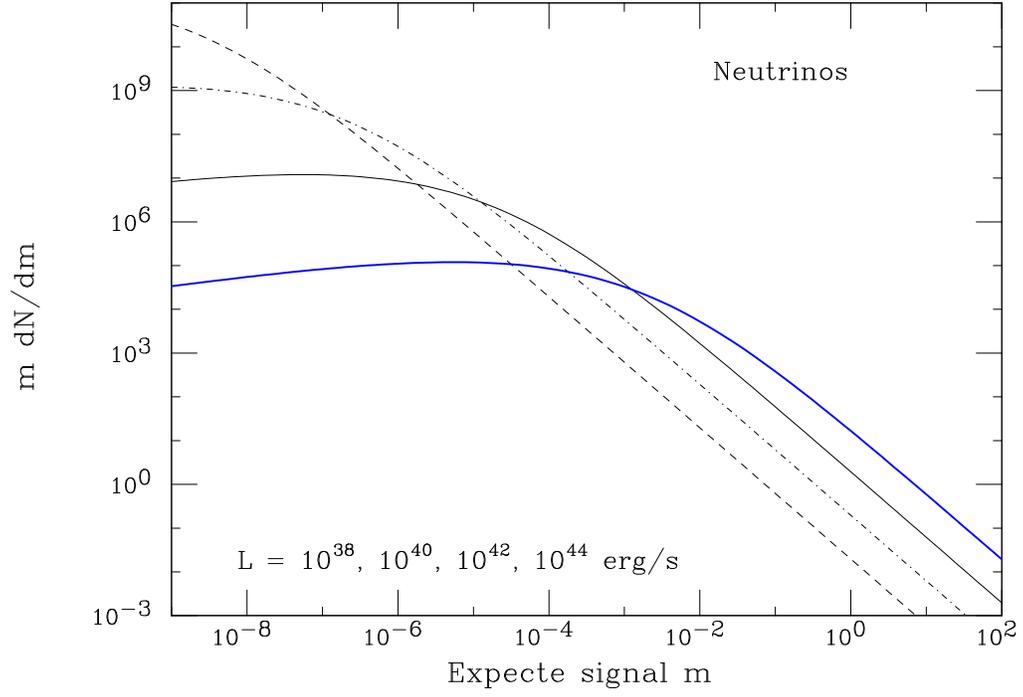,angle=90,width=13.5cm}}
\caption {\footnotesize 
Number of sources with an  expected  signal $m$
(plotted in the form $ m \; dN/dm$   versus  $m$).
  The distributions
are  calculated for the sam,  power density ${\cal L}_\nu$,
assuming that all sources   have the same 
luminosity $L$ and  emit with a 
a power law  spectrum of slope $\alpha = 2$,  
The different  curves   differ  only because of the value  of $L$.
\label{fig:nu4}  }
\end{figure}

\begin{figure} [hbt]
\centerline{\psfig{figure=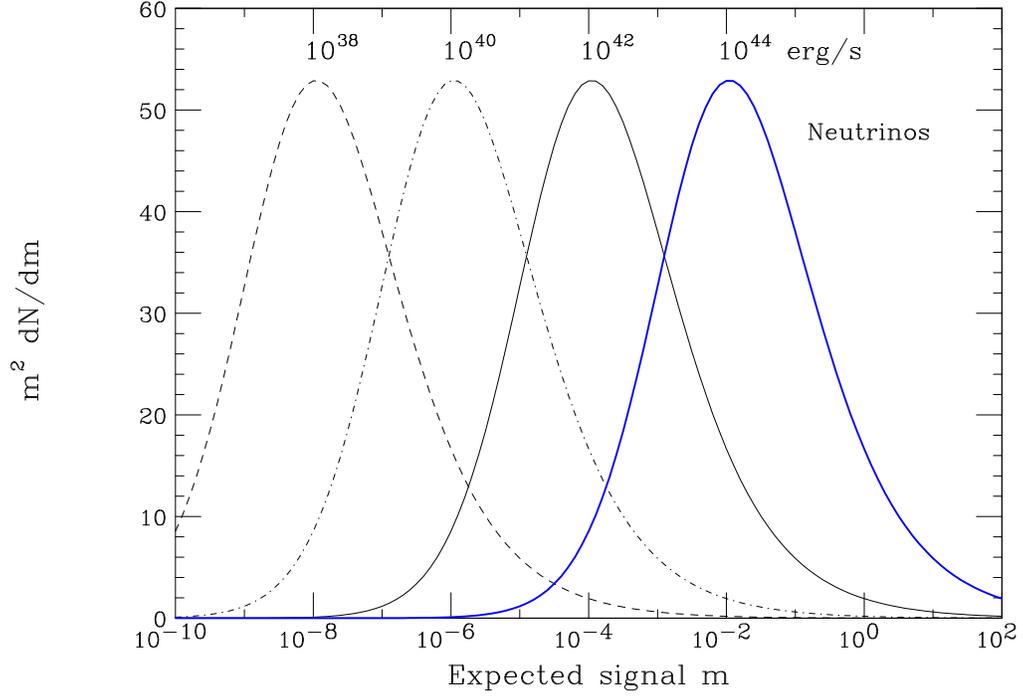,angle=90,width=13.5cm}}
\caption {\footnotesize 
As in  figure~\protect\ref{fig:nu4},    but plotted 
in the form $ m^2 \; dN/dm$   versus  $m$.
The area  under each curve is proportional
to the  total  number of  detected neutrino events.
\label{fig:nu5}  }
\end{figure}

For a qualitative understanding of the role of the function
$G_{m , \Phi}(x)$ one  can look at  figures~\ref{fig:nu4}
and~\ref{fig:nu5} that  show the  distribution of
expected signal  
 $m = \Phi \, A \, t$  for an  isotropic 
neutrino detector.
Neutrinos  are emitted  by an  ensemble of identical sources
of  luminosity $L_0$  emitting 
with a  power law spectrum  of  slope $\alpha = 2$;
the cosmological parameters  are $\Omega_m = 1$ and  $\Omega_\Lambda = 0$.
The   injection power density ${\cal L}$ 
is  taken to the same  value, but  different calculations assume
different values for the luminosity   $L_0$ of  the individual sources.
Changing $L_0$ leaves the  total  signal
unchanged, but  generates  very different  
 distributions $dN/dm$.
Figure~\ref{fig:nu5} shows the  distribution
$dN/dm$ is the form $m^2 \, dN/dm$ versus $m$  with 
a logarithmic scale for $m$.  In this  way  the area
under the cuve is proportional to inclusive signal 
and the shape of  the curve is   identical to the
shape of the curve $x^3 \; G_{\Phi,m}(x)$.  
It is  simple to see that   the power density
of the sources  change the absolute normalization of  the curve,
the luminosity $L_0$  ``shifts''  the   distribution to higher or
smaller   signals, with the overall shape  determined
by  the scaling function $G_{\Phi}(x)$ that is 
calculable from  the  particle  propagation properties,
the cosmological parameters, and the shape of the luminosity function.

\section{Interpretation of the  data on  UHECR clustering}
It is possible to use the  instruments   developed 
in  the previous sections to  analyse the
results  obtained by   UHECR detectors.
As an  example we will  discuss 
 the  results of the Auger detector  
for the  first  integrated exposure $A_\Omega \; t = 9000$~(Km$^2$~yr~sr)
\cite{Cronin:2007zz,Abraham:2007si}.
In this  exposure  the
detector  has  observed   27 events
with  energy above $E_{\rm th}  \simeq  6 \times 10^{19}$~eV.
From this  event rate it is possible
to estimate the power density  $ {\cal L}_p (E_{\rm th})$ of the 
emission in the energy
interval [$E_{\rm th}, \; 10\, E_{\rm th}]$ 
making the assumptions  that the particles are protons, and 
using equation (\ref{eq:nev})
that  can rewritten  as:
\begin{equation}
N_{\rm events} = 
17.2 ~ 
\left ( \frac{ {\cal L}_p (E_{\rm th})}{10^{36}~{\rm erg}/{\rm s}\;{\rm Mpc}^3} \right )
~\left ( \frac{A_\Omega \; t }{9000~{\rm Km}^2~{\rm yr}~{\rm sr}} \right )
~\left ( \frac{\zeta_\Phi (E_{\rm th}) }{0.037} \right )
~\left ( \frac{\eta}{0.434} \right )
\label{eq:nev1}
\end{equation}
From the observed event rate one can   estimate the power  density is:
\begin{equation}
 {\cal L}_p (E_{\rm th})  \simeq  1.6 \times 10^{36}
~\left (\frac{E_{\rm th}}{60~{\rm EeV}} \right )
~\left ( \frac{0.037}{\zeta_\Phi(E_{\rm th} )} \right )
~\left ( \frac{0.434}{\eta} \right )
~\frac{ {\rm erg}}{{\rm s}~{\rm Mpc}^3}
\label{eq:pow0}
\end{equation}
The error  on  the determination of the
power density  is  dominated by systematic effects
associated   with uncertainties in 
the  determination of the  energy scale of  the  measurement.
An error  in the energy scale  affects   obviously 
the   value  of $E_{\rm th}$,  in addition,  the value
of the shape  factor $\zeta_\Phi(E_{\rm th})$  is  a very steep
function  of energy. 
For $E \sim E_{\rm th}$ this  energy dependence  can be approximated
with 
$\zeta_\Phi (E) \propto E^{-2.3}$.
Therefore one can  make estimate the uncertainty
on the power density as:
\begin{equation}
  \frac{\Delta  {\cal L}_p (E_{\rm th})}{ {\cal L}_p (E_{\rm th})} \simeq
0.20~({\rm stat}) + 3.3 ~\left (\frac{\Delta E}{E} \right )_{\rm scale}
\end{equation}
The shape  of the CR energy spectrum 
is  also important.   It determines
the quantity  $\eta$  that  connects  the integral emission  
and the  power  density 
  ($Q_{\rm incl} (E_{\rm th})  = {\cal L}(E_{\rm th}) \; \eta/E_{\rm th}$), and also
 the   precise value of the  shape factor  $\zeta_\Phi(E_{\rm th})$.
 If the  spectrum is a power law with 
 a slope that varies in the  interval $\alpha \in [2,2.6]$
 then  $\eta$   (see equation (\ref{eq:eta}))   varies in the
 interval  $\eta \in [(\ln 10)^{-1} = 0.434, ~0.500]$,
with  a steeper  spectrum   requiring   less  power for the same  flux.
 For the same  range of  slope,
 the   shape factor $\zeta_\Phi(E_{\rm th})$  
(for $E_{\rm th}  = 60$~EeV) 
 varies in the  interval  $\zeta_\Phi \in [0.037, 0.032]$, in this case 
a  steeper  spectrum  has a smaller  horizon  and therefore  requires
more  power to  generate the same flux.
The  effects of  changing the slope $\alpha$
go  in opposite directions for $\eta$  and $\zeta_\Phi$  and therefore
in first approximation cancel each other in the estimate of the 
power  density.

The  determination of the shape of the extragalactic flux 
of  CR   requires  the disentangling  of the  galactic  component.
A crucial problem is the determination of the  transition energy 
where the  two components  (galactic and extragalactic)  are  equal. 
If the  transition energy corresponds to the ``ankle''
at $E \simeq 10^{19}$~eV, the slope of the spectrum is  only
poorly determined experimentally.  
Theoretical  prejudice
has  traditionally   favored  a  flat slope $\alpha \simeq 2$.
If the extragalactic component 
dominates down to much lower  energy  as advocated
first by Berezinsky and his collaborators  \cite{Berezinsky:2005cq},
the   slope $\alpha$   is    better determined
and is significantly larger than 2.
If the evolution of the sources  is  negligible,
one estimates  $\alpha \simeq  2.6$,
 if  the source evolution has  a behaviour   
similar   to the Stellar Formation Rate (or the AGN  luminosity function)
one obtains $\alpha \simeq 2.4$.  

If uncertainties in the spectral  shape  are   not  very significant
in the estimate of  the power  density in the energy decade
above 60~EeV, they have a   very dramatic  effect if one attempts
to   estimate  a  total CR power  density  integrated over all
energies.   Extrapolating the power  low  behaviour
to the  interval  $\{E_{\rm min},E_{\rm max}\}$  one    can easily  compute:
\begin{equation}
{\cal L}_p [E_{\rm min}, E_{\rm max} ]
 =  {\cal L}_p [E_{\rm th}, 10\, E_{\rm th}] ~ \frac{\eta}{(\alpha -2)}~
\left [ 
\left ( \frac{E_{\rm min}}{E_{\rm th}} \right )^{-\alpha +2}
-
\left ( \frac{E_{\rm max}}{E_{\rm th}} \right )^{-\alpha +2}
\right ]
\end{equation}
In the limit of $\alpha = 2$ this  becomes:
\begin{equation}
{\cal L}_p [E_{\rm min}, E_{\rm max}]
 =  {\cal L}_p [E_{\rm th}, 10\, E_{\rm th}] ~ 
\frac{\ln (E_{\rm max}/E_{\rm min}) }{\ln 10}
\end{equation}
with  a weak, logarithmic  dependence on the energy limits.
If  $\alpha$ is large the low  energy extension of the spectrum
is  problematic  because the energy requirement
grows  rapidy  with   decreasing $E_{\rm min}$, in good approximation
$\propto (E_{\rm min}/E_{\rm th})^{-\alpha +2}$.
As an illustration the extension 
as  an unbroken power law
of the extragalactic  CR spectrum 
down to  $E_{\rm min} \simeq 10^{12}$~eV  
requires  a  power  approximately 9   (2100) times  larger than
the estimate of   equation (\ref{eq:pow0}) if one assumes 
a slope  $\alpha =2$  ($\alpha =2.4$).

\begin{figure} [hbt]
\centerline{\psfig{figure=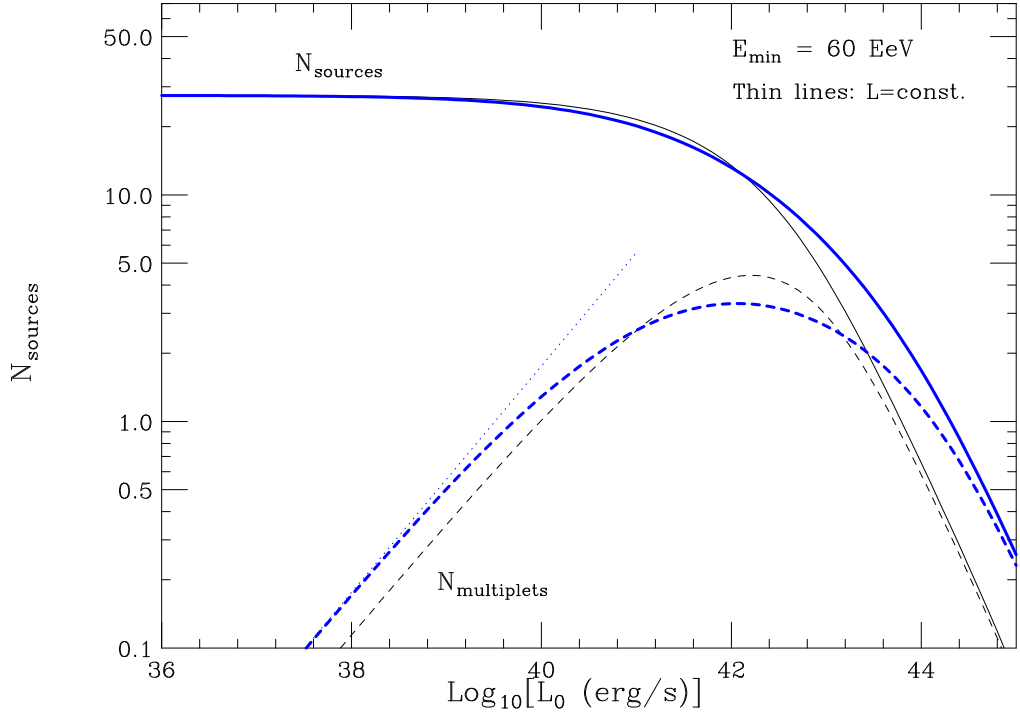,angle=90,width=13.5cm}}
\caption {\footnotesize 
Expected total  number of detected sources (solid lines) and  
number of sources  with multiplicity $k \ge 2$  (dashed lines) plotted as 
a function of  the median luminosity of the sources.
The  calculation assumes 
a threshold energy  $E_{\rm min} \simeq 6 \times 10^{19}$~eV
and a total exposure of 9000~(Km$^2$~yr~sr) for the  Auger detector.
The source power  density is set to ${\cal L}_p \simeq 1.6 \times
10^{36}$~erg/(s~Mpc$^3$) to reproduce  the observed event rate.
The thin (thick) curves assumed a  fixed  luminosity (the luminosity function
of  equation (\protect\ref{eq:lum_shape}).
The dotted  line is $\propto \sqrt{L_0}$.
\label{fig:auger1}  }
\end{figure}

\vspace{0.4 cm}
The clustering of the  events detected
above the threshold  energy $E_{\rm th} = 6 \times 10^{19}$~eV
can  give  us information about how the power
density  (\ref{eq:pow0})  is  subdivided in   individual sources
{\em if}   we assume that  the propagation of protons
of these energies  for   distances of the order
of the corresponding horizon is  approximately linear.
The   general   method to  compute the expected number of 
clusters  with  multiplicity  $N_k$  
has  been discussed  before, and 
involves  the calculation of the  distribution
of  expected  signal  $dN_s/dm$, and the folding 
with a Poissonian  probability.
The results of a    numerical  calculation of  
$N_{\rm multi}$   (the number of clusters
with multiplity $k \ge 2$) and of  detected sources
as  a  function of the median luminosity $L_0$ 
is  shown in  figure~\ref{fig:auger1}.
The  calculation  is  performed 
for    the  value of the source power density
that  reproduces  the inclusive rate, and using  two
models for  the  luminosity function:
a  fixed luminosity,   and
the broad luminosity distribution  given in equation (\ref{eq:lum_shape})
and shown in figure~\ref{fig:lum_shape}.

The  behaviour of the  curves  $N_{\rm sources} (L_0)$ 
is simple to understand  qualitatively.
For small $L_0$  the     cosmic ray  signal is  given by the contribution
of many  faint sources, each  contributing  a small  fraction of an  event 
and one has $N_{\rm sources} \simeq N_{\rm  events}$, with a  negligible
probability of observing multiplets.  Increasing
$L_0$  the number of  detected  sources  
(for a  fixed  power density)   decreases  monotonically  as the
number  density of sources   decreases, while the volume
contained inside the horizon   remains  fixed.

The  behaviour of $N_{\rm multi}(L_0)$ as a function of  $L_0$   shows
a maximum for  $L_0 = L_0^*  \sim 10^{42}$~erg/s. 
For  a  median luminosity $L_0 \ll  L_0^*$  
the   number of multiplets  grows   $\propto L_0^{1/2}$
according to equation:
\begin{eqnarray}
N_{\rm multi} &= & 
\sqrt{\pi} ~\left ( \frac{ 4 \, \pi }{3} \right )~
 {\cal L}_p (E_{\rm th}) ~\sqrt{L_0} ~
~\left ( \frac{ \eta \; A_\Omega \, t }
{ 16 \, \pi^2 \; E_{\rm min}} \right )^{3/2}
~ \langle  a^{3/2} \rangle
~ \langle  y^{3/2} \rangle
\nonumber \\
& ~ & \nonumber \\
& \simeq &
0.74 ~ 
\left ( \frac{{\cal L} (0)}{10^{36}\;{\rm erg}/{\rm s}\;{\rm Mpc}^3} \right )
~\sqrt{\frac{L_0}{10^{40}~{\rm erg}/{\rm s}}}
~\left ( \frac{A_\Omega \; t }{9000~{\rm Km}^2\;{\rm yr}\;{\rm sr}} \right )^{3/2}
\langle y^{3/2} \rangle
\label{eq:multiplets}
\end{eqnarray}
The  behaviour  $N_{\rm multi} \propto  \sqrt{L_0}$
has a very simple  explanation.  The  radius   of the sphere
within  which a source of  luminosity $L_0$  is  sufficiently    bright
to  generate a multiplet   grows  approximately 
 $\propto \sqrt{L_0}$, accordingly the volume   that contains
multiplets grows $\propto L_0^{3/2}$; on the other hand,  for
a  fixed luminosity density  the  number  density of the sources
is $n_s \propto L_0^{-1}$,  with the combined  result
$N_{\rm multi} \propto  L_0^{3/2} ~L_0^{-1} \propto L_0^{1/2}$.
This  behaviour    stops  when the  radius  of detection
of  a  source   becomes   comparable with the ``horizon  radius''  
of order  $(R_0 \,  \zeta_\Phi)$.  Additional  increase in $L_0$   decreases
the number   of the multiplets, since the detection  volume remains
approximately constant  while the  source  number  density decreases.
The coefficient of proportionality between
$N_{\rm multi}$ and $\sqrt{L_0}$  depends  linearly on the 
factors  $\langle  a^{3/2} \rangle$  
and
 $\langle  y^{3/2} \rangle$.
The first, defined in equation  (\ref{eq:aomega}), depends
on the detector  geometry 
and event  selection criteria.
For the  Auger detector location and the 
cut on zenith angle $\theta < 60^\circ$
one has  $\langle  a^{3/2} \rangle \simeq 1.26$.
The factor $ \langle  y^{3/2} \rangle = 1$ is  
unity if all sources  have the same  luminosity;
for the broad  luminosity distribution  (\ref{eq:lum_shape})  
$ \langle  y^{3/2} \rangle \simeq 1.64$.
It is  intuitive  that a  broad  luminosity distribution (with large
$\langle  y^{3/2} \rangle$)  yields a larger number
of multiplets  than a   narrow one  thanks to the contribution
of  high luminosity  sources  in the tail of the  distribution.

In the  regime of  validity of  equation (\ref{eq:multiplets})
It is possible to obtain  $L_0$  with  a simple, closed form  expression
as  a  function of $N_{\rm  events}$ and $N_{\rm multi}$ 
solving  the system  composed of  equations  (\ref{eq:nev}) and
(\ref{eq:multiplets}):
\begin{eqnarray}
L_0 & \simeq &  144 \, \pi ~ 
\left  ( 
\frac{N_{\rm multi}}{N_{\rm events}} \right )^2
~\frac{E_{\rm th}  ~[R_0 \; \zeta_\Phi(E_{\rm th})]^2}{\eta ~(A_\Omega \; t)}
~ \frac{1}{\langle a^{3/2} \rangle^2} 
~ \frac{1}{\langle y^{3/2} \rangle^2} 
\end{eqnarray}
Substituting the  number of events observed in Auger one finds:
 \begin{equation}
 L_0 \simeq 7.5 \times 10^{39}~
 \frac{(N_{\rm multi})^2}{ \langle y^{3/2} \rangle^2}
 ~\left ( \frac{E_{\rm th}}{60~{\rm EeV}} \right )~
 ~\left ( \frac{\zeta_\Phi(E_{\rm th})}{0.037} \right )^2~
 ~\frac{\rm erg}{\rm s}
 \label{eq:L0auger}
 \end{equation}
Note that  in the  regime   of validity of 
this  equation 
the number of  multiplets  grows
with time as: $N_{\rm multi} \propto t^{3/2}$, while
 the number  of  events is  
$N_{\rm events} \propto t$, therefore  
the combination [$(N_{\rm multi}/N_{\rm events})^2 \; t^{-1}$] is 
a constant.
The   behaviour 
$L_0 \propto (N_{\rm multi}/N_{\rm events})^2$ 
loses is  validity for $L_0 \gtrsim 10^{41}$~erg/s, when  sources
not too distant from the energy loss horizon    begin to be   visible.

 The  question of how   the  27  highest energy
 events  detected by Auger can be subdivided  into
 the contribution of different sources  has not  a simple  answer, and is
 intimately related to the problem of the 
 the deviations   of the particles  during their propagation
 and the  size of the 
 extragalactic  magnetic fields.

 The  ``nominal interpretation''  of the Auger  events,
 suggests that  particles  with $E \gtrsim 60$~EeV  are  deviated by
 approximately 3 degrees, and therefore that the clustering
 of events should be  searched for  using a correlation  cone
 of approximately this size.
 If one  tentatively accepts  this  conclusion, 
defining a  cluster
 as  a  set of events  with angular  separation  less than a $3^\circ$,
 one finds  2 doublets  and correspondingly 23  singlets.  
Increasing the size of the correlation cone to 5$^\circ$ one finds
one triplet  and   three doublets,  while   the 18  remaining events  are  singlets.
These  numbers  should  be  corrected  for the  effect
of random coincidences   that in turn depends on the
assumptions  about the larger  scale  angular  distribution of the  flux. 
Taken at  face value  these  results, together with the 
the  calculations shown in figure~\ref{fig:auger1}  
suggest   an allowed   range  for the luminosity $L_0$:
\begin{equation}
 10^{39} ~\frac{\rm erg}{\rm s} \lesssim  L_0 \lesssim 10^{41.5} ~\frac{\rm erg}{\rm s}
\label{eq:lum_estimate}
\end{equation}
The lower  limit  follows  from the fact that 
(at a modest confidence level)  some  clustering 
is  observed. 
The upper limit  follows from the fact that most
of the events are actually in singlets, and that the total number
of  objects that contribute to the signal is  larger  than $\simeq 20$.

In fact  the Auger data suggest that at least one source
of UHECR has  been established with a 
very high  degree of confidence. 
 The position  of  the  very  close Active Galactic Nucleus
 Cen~A  coincides with  the most prominent  of all  tentative
clustering structures  (the triplet 
found for a 5$^\circ$ aperture cone).
Considering  that  Cen~A,
on   the basis of its  properties and closeness to the Earth,
has  been  proposed as one of the most likely  sources of UHECR  
this is a striking  result.
The detection of Cen~A is  an remarkable result of  historic importance.
It is the  first imaging  of an astrophysical object 
with   charged particles and,    4  centuries after the  
invention of the  optical telescope and 20 years  after the detection
of  SN1987A  with neutrinos,  marks the birth of 
a new astronomy. 

The  clarification of the size and structure of
the   excess of  events from the Cen~A direction
is  very important to evaluate the perspectives of the field of CR astronomy.
In fact it is possible  that  Cen~A  
accounts for a    number of events larger than  2 or 3.
 For  example a  10$^\circ$  ($20^\circ$) cone
around the  direction of Cen~A  contains  4 (9)  events, suggesting that
the excess from this  direction is  larger and  more  extended.
It   should  be stressed that Cen~A is   remarkably close to  us
with a  distance of only 3.5~pc.  Since the deviations
of a particle  during propagation in intergalactic space
scales   as $\sqrt{d}$ (with $d$ the sorce distance)   a  signal with an
angular size of several degrees from  Cen~A suggest  that  
astronomy from sources  at larger distances will be  very difficult
and  require  higher  energies.
The result is also in possible conflict  with  the correlation cone
of 3.2  degrees  found  between the direction of  the highest energy cosmic
ray  and the  position of AGN with $z \le 0.017$  found in \cite{Cronin:2007zz}
since the  distance of these AGN is   typically
 20 times larger, and the expected  deviation 
of particles produced in sources is expected to be 4 times larger.

There are several solution to this problem:
the apparent large angular size  of the event excess in the direction
of Cen~A  could be  due to   the contributions of other sources
in the same   portion of the sky;  the large  angular deviations  could
be   the  effect of strong magnetic fields  in the vicinity 
of the source and not in intergalactic space
(but of course this  should  then be  expected 
for a significant  fraction  of all  the other sources);  the 
angular  spread of the signal could be associated
with the  size of the emission region from Cen~A that  has  jets
extending  several degrees away from its nucleus.
It could however also mean that the   intergalactic  magnetic  fields
are  stronger than  expected, and that  
cosmic  ray extragalactic astronomy is  only 
 possible only for very close  sources  or very high energy.

In any case, the power  emitted in cosmic  ray  from  CenA  
is straightforward to compute. In fact the  source
is so close  to us   that   the cosmological  and energy
loss corrections are  small
for $E \sim  6 \times 10^{19}$~eV.  
The luminosity in the energy decade
above $E_{\rm min}$ can be estimated as:
\begin{equation} 
L_{\rm CenA} \simeq 1.5 \times 10^{39}~
\left (\frac{N_{\rm CenA}}{2}\right )
~\left (\frac {d} {3.5~{\rm pc}}\right )^2
~\frac{\rm erg}{\rm s}
\label{eq:lumcena}
\end{equation}
(where  $N_{\rm CenA}$ is the  number of events arriving from Cen~A).
This  result  is in reasonable agreement 
(at least in order of magnitude) with the estimates
(\ref{eq:L0auger}) and (\ref{eq:lum_estimate}), especially allowing
for a broad  luminosity distribution.

\begin{figure} [hbt]
\centerline{\psfig{figure=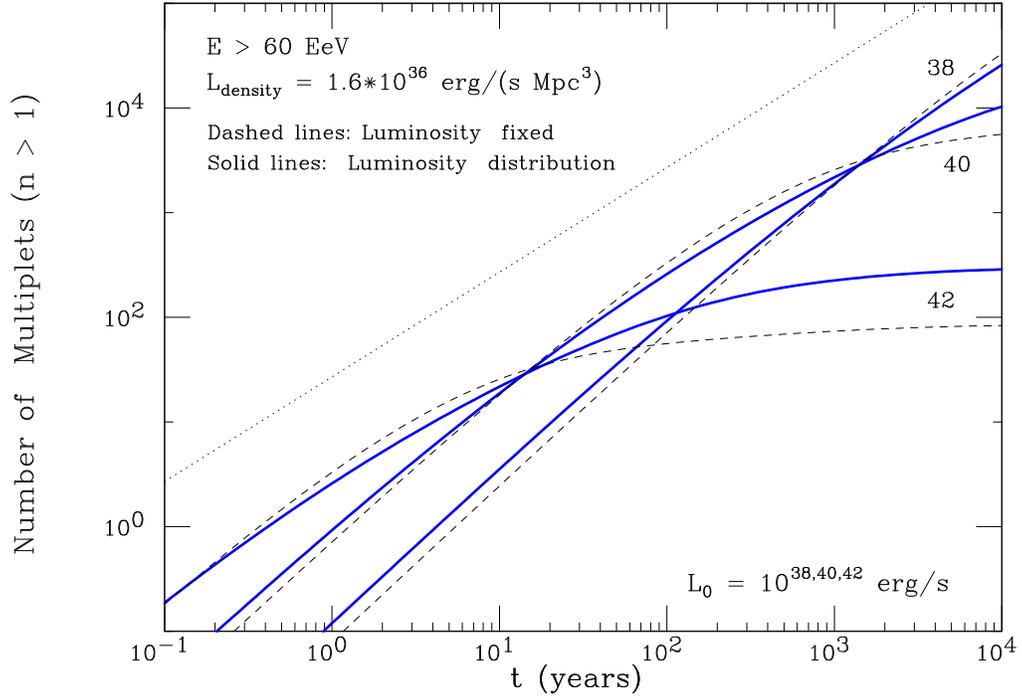,angle=90,width=13.5cm}}
\caption {\footnotesize 
Expected  number of  proton sources  detectable in Auger with 
multiplicity  $n \ge 2$ 
(for $E_{\rm min} = 6 \times 10^{19}$~eV)
 plotted as a function of detector  live time.
The power density of the sources is  
${\cal L}_p = 1.6 \times 10^{36}$~erg/(s~Mpc$^3$).
to  reproduce   the observed inclusive event rate
(the  dotted line  gives the total number of events as a function
of time).
The different curves are  calculated 
assuming that all sources  
have the same  spectral shape 
(a power law of slope  $\alpha = 2.0$) 
and   3 possible values of
the median luminosity:
($L_0 = 10^{38}$,
$10^{40}$ and
$10^{42}$ erg/s).
The   dashed  line are calculated  with a fixed
luminosity, while the  solid lines
are the  result  using the 
luminosity  function of equation
(\protect\ref{eq:lum_shape})  and shown in 
fig.~\protect\ref{fig:lum_shape}.
\label{fig:auger2}  }
\end{figure}

It is  interesting to  make a  prediction
on how the number of detected  clusters  will scale
with  additional exposure.
An illustration of this  problem is in~fig.~\ref{fig:auger2}
that shows how the  number of clusters with multiplicity
$n \ge 2$ grows with exposure for the Auger detector.
If the median luminosity of the  sources  is small
the number of  clusters  grows $\propto (A \, t)^{3/2}$,  faster than
the linear  growth of the inclusive  rate.  Eventually this  growth  must 
slow  down and finally stop  when the  exposure is  sufficient 
to  observe as  clusters of events sources  
close to  the energy loss horizon.  The asymptotic number of 
detectable  sources is clearly sensitive to the form of the luminosity
function,  and to the  density of weak sources  with luminosity
much smaller than the median  luminosity $L_0$.

\section{Extragalactic  neutrinos}
The  existence  of  the  extragalactic  cosmic  rays  is   the
best reason to  predict also a measurable 
flux of extragalactic neutrinos. In  fact, the CR sources
must  unavoidably  also generate  neutrinos,  
because a  fraction of the relativistic 
hadrons they produce   will interact  with matter or  radiation fields
inside or near the  source, 
producing weakly decaying  particles (pions and kaons) 
that generate  neutrinos.
The power  density and luminosity function of the 
neutrino extragalactic sources  are
therefore  related  to the   analogous  quantities for   cosmic  ray emission.
The relation between the   CR and  the $\nu$  emissions
is  very uncertain and  depends on the structure of the  sources.
The  motivation for the simultaneous measurement 
of both  cosmic rays  and neutrinos is  precisely the 
understanding  of the nature and  structure of their common sources.

In the following we want to discuss the observations 
of neutrinos  using the   detection
of  $\nu$--induced    muons in ``Km$^3$'' type neutrino telescopes
like IceCube  at the South Pole and Km3Net  in the Mediterranean sea.
We will use a   simplified   model for the  effective  area  
of a neutrino  detector  that is   discussed in appendix~\ref{sec:app_nu}. 
It can  be useful to think   about the signal of
neutrino  induced  muons  as  a  measurement of
the integral  neutrino  flux  above    the threshold
$E_\nu^{\rm min}\simeq 1$~TeV  using an effective  area
that is the geometric  area of the  detector  times an energy averaged
neutrino to muon  conversion efficiency 
of  order  $\langle \varepsilon_{\nu \to \mu}\rangle \sim 3 \times 10^{-6}$.
The exact value  of  $\langle \varepsilon_{\nu \to \mu}\rangle$  depends on the  shape of the neutrino spectrum,
the energy threshold for muon detection  and the average zenith angle
of the source considered.
Taking into account the small   conversion efficiency
the effective area of a Km$^3$  neutrino detector  is of order
of only  few  square  meters,  many orders of magnitude smaller  that
the effective  area of  a telescope  like
Auger  (with a surface of approximately 3000~Km$^2$).
It is  however interesting to    note that, in some sense,
the sensitivities of    detectors like Auger and  
IceCube are   comparable.
This  can be illustrated  with the following example,
if   we have one source that emits 
protons  and  neutrinos  with spectra that are
approximately equal in both  shape and  absolute normalization,
and the  spectra  can be reasonably well approximated
as power  law   of slope
$\alpha \simeq 2$, the the  source 
(if in the field of view of the detectors) 
will leave event rates in
ultra high energy showers, and in  neutrino induced muons
are of the same order of magnitude.
This can be easily understood qualitatively 
observing that  energy threshold shower detection
is  of order  $\sim 10^{20}$~eV,
while the neutrino  telescopes  is sensitive to neutrinos
with $E_{\nu} \gtrsim 10^{12}$~eV. The integral fluxes
above  these energy thresholds are 
(for a spectrum of slope  $\alpha \simeq 2$) in a
ratio  $10^{-8}$, but also the effective areas
differ  by approximately  the same  factor, resulting
in approximately the same    event rates.

\begin{figure} [hbt]
\centerline{\psfig{figure=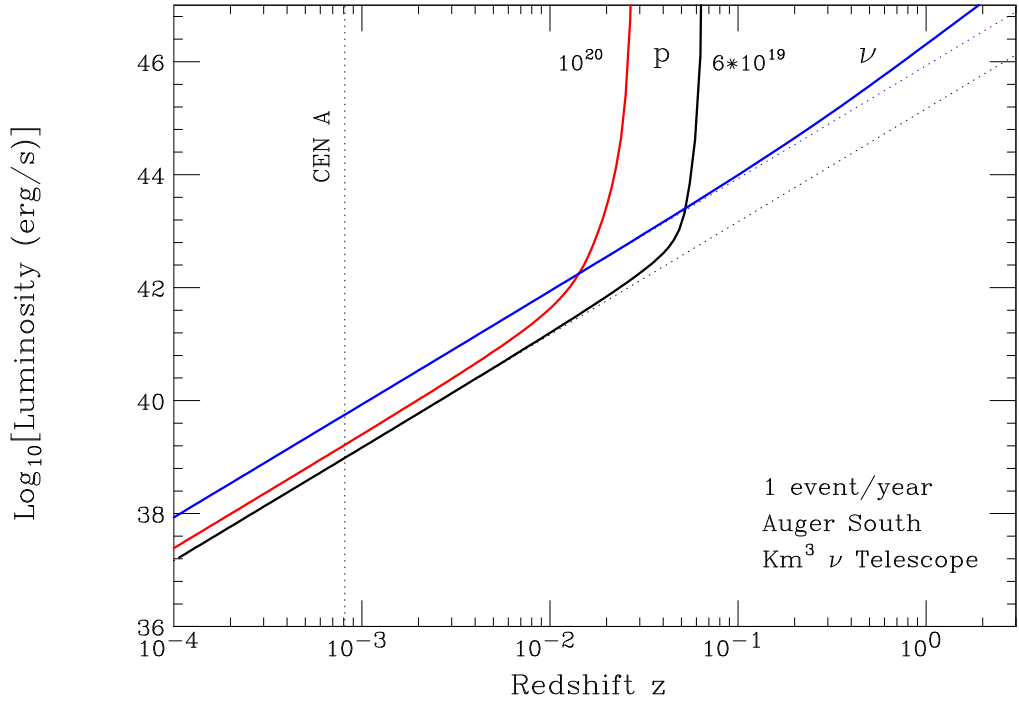,angle=90,width=13.5cm}}
\caption {\footnotesize 
The lines describe the curves in the
\{$z$, $L$\}  (luminosity, distance)  plane that correspond to  a rate of one event/year
from a source. The luminosity must be understood
as  luminosity per energy decade.
The different curves are 
for the detection of proton  showers 
with threshold energy  $E_{\rm min} = 60$ and 100~EeV 
in Auger  (assuming  a source  declination $\delta \simeq -30\circ$),
and for the number of neutrino  induced muons 
in IceCube (for a  source with $\delta > 0$).
The dotted lines are $\propto z^{2}$.
\label{fig:lumz}  }
\end{figure}

This  discussion is made  more quantitative in figure~\ref{fig:lumz}
 that show  the curves  in the plane
$\{z,L\}$  (distance, luminosity)
that corresponds  to  one  event/year in  Auger 
(for a source  places at $\delta \simeq -30^\circ$) and in
a Km$^3$  telescope of   the  IceCube type.
The luminosity must be understood as the luminosity
per energy decade, and the source
spectrum is a power law with slope  $\alpha = 2$.
For  very close   sources,  the 
lines that correspond to a constant    event rate
have the simple form  $L(z) z^{2} \simeq {\rm const}$.
For these close sources, the  neutrino event rate
is  a factor  approximately~5 (3) times  smaller 
than the shower  rate  with a  threshold
$E = 6\times 10^{19}$~eV ($E =  10^{20}$~eV).
On the other hand the proton  source  become  
effectively invisible when they are behind
an energy loss  horizon that  is a  rapidly varying
function of energy, while  the entire  universe
is  visible with neutrinos.

It is  interesting to perform  also for 
extragalactic neutrinos
the same   study done for protons,
and  discuss  what  fraction of the
inclusive  signal 
can  be resolved as   the contribution of individual sources.

With respect to the  study of UHECR 
there are a several  obvious differences.
The first one  is that   neutrinos  travel  in straight lines
and therefore the problem of the impact of extragalactic
magnetic  field  is  entirely absent. 
The angular size of a  cluster of $\nu$--induced muons is    controled
by the muon--neutrino angle,  that is    calculable
(the angular dimension of the signal  has  a weak  but non
entirely negligible  dependence on the 
shaope of the neutrino  spectrum). 

A second  difference is the  presence of the atmospheric  neutrino  background
that  constitute an important complication. The total number
of  neutrino events from  astrophysical  sources
can only  be obtained with a non trivial  analysis
that  requires the estimate and  subtraction 
of  the background. 

An additional  problem is that  much 
(but not all) of the  information about the   shape
of the neutrino energy spectrum is lost. 
The  detected muon
energy can  only be measured  with  poor  resolution, and  this
energy is  only poorly  correlated with the parent neutrino energy.

Finally, neutrinos  can  travel 
without interacting across a Hubble distance, and it is therefore
expected that the   fraction of the inclusive signal
that can be  resolved and  associated 
to individual sources, is going to be  much smaller
than for UHECR    that  can reach us   only from a much smaller   volume
determined  by the  energy loss length.

\vspace{0.35 cm}
At present  there is   only an upper limits on the possible
flux of extragalactic  neutrinos. 
The most  stringent  one has been obtained by the 
Amanda detector at the South Pole \cite{Achterberg:2007qp},
and is  expressed 
in reference to  an isotropic  neutrino  spectrum
of form  $\phi_\nu = K_\nu~E^{-2}$.  The 90\% C.L.
upper limit on the constant $K_{\nu_\mu}$ (relevant
for the combination $\nu_\mu + \overline{\nu}_\mu$) is
\begin{equation}
K_{\nu_\mu} \equiv E^2 ~ \phi_{\nu_\mu }  \le
7.4 \times 10^{-8}
~{\rm GeV}/({\rm  cm}^2 \; {\rm s} \; {\rm sr})
\end{equation}
The  quantity  $K_{\nu_\mu}$ can be 
expressed in terms of the power density
of the neutrino  sources at the present epoch using:
\begin{equation}
3 ~K_{\nu_\mu} = \frac{R_0}{4 \, \pi} \; \frac{ {\cal L}_\nu^0}{\ln 10} ~\zeta_\Phi
\end{equation}
where the factor of 3  takes into  account
the expected contribution of all other  neutrino flavors
(see  \cite{Lipari:2007su} for a critical discussion).
The upper  limit on  the   power density of the extragalactic
neutrino sources is:
\begin{equation}
{\cal L}_\nu^0 \lesssim
 1.04 \times 10^{37} ~
~\left (\frac{2.2}{\zeta_\Phi} \right )
~  \frac{\rm erg}{{\rm Mpc}^3 \; {\rm s}} 
\end{equation}
(The  estimate $\zeta_\Phi \simeq 2.2$ includes an important  redshift dependence of
the source power density).
It can be  useful  to  translate this upper limit on  the 
extragalactic neutrino flux as  a 
maximum number of events generated  from extragalactic neutrinos.
The total  number of extragalactic  neutrino  events
can be calculated   using  equations (\ref{eq:nev}), and depends
on the power  density of the neutrino   events:
\begin{eqnarray}
N_{\rm events} & \simeq & 
 1270
~\left (  \frac{ {\cal L}_\nu^0}
            { 10^{37} ~{\rm erg}/({\rm  Mpc}^3{\rm s})} \right )
~\left (  \frac{\zeta_\Phi}{ 2.2}  \right )
~\left (  \frac{\varepsilon_{\nu \to \mu}}{ 3 \times 10^{-6}}  \right )
~\left (  \frac{\eta}{0.434}  \right )
~\left (  \frac{A_0 ~ t}{ {\rm km}^2~{\rm  yr} } \right )
~~~~
\end{eqnarray}

\begin{figure} [hbt]
\centerline{\psfig{figure=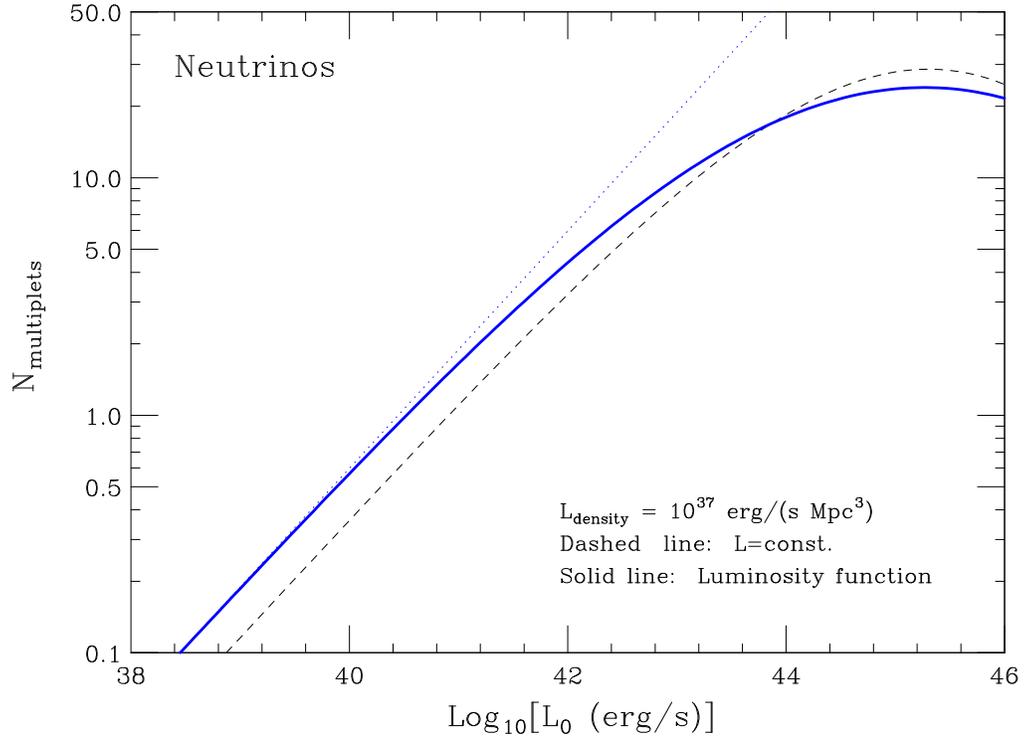,angle=90,width=13.5cm}}
\caption {\footnotesize 
Expected  number of  neutrino  event  clusters
with $k \ge 2$  plotted as a function of the median luminosity
$L_0$  of the  neutrino sources.
The calculation is performed  with a neutrino  power  density 
${\cal L}_\nu^0 = 10^{37}$~erg/(s~Mpc$^3$), and assuming
no  cosmological  evolution of the sources.
The solid  lines  assume   the luminosity function
of equation (\protect\ref{eq:lum_shape}), the dashed
lines a fixed  value $L=L_0$.
\label{fig:nu_mult}  }
\end{figure}

 A  calculation of the number of sources and of   the number of
multiplets  as  a function of the median luminosity $L_0$ is  shown
in fig.~\ref{fig:nu_mult}.
The calculation is  performed  for a power density of the neutrino
sources  ${\cal L}_\nu^0 = 10^{37}$~erg/(s~Mpc$^3$)   just  below
the  existing  constraints.
For  $L_0$ not too large  the expected  number  of multiplets
is  proportional to $\sqrt{L_0}$ 
according to the approximationo of equation
(\ref{eq:multiplets})  that can be recast in  numerical  form as:
\begin{eqnarray}
N_{\rm multi} 
& \simeq & 3.7
~\left (  \frac{ {\cal L}_\nu^0}
            { 10^{37} ~{\rm erg}/({\rm  Mpc}^3{\rm s})} \right )
~\sqrt{\frac{L_0}
            { 10^{42} ~{\rm erg}/{\rm s}} }
~\left (  \frac{A_0 ~ t}{ {\rm km}^2~{\rm  yr} } \right )^{3/2}
~\langle y^{3/2} \rangle
\label{eq:mult_nu}
\end{eqnarray}
In   equation (\ref{eq:mult_nu}) 
the quantity $\langle a^{3/2} \rangle $   has been  estimated  as
$\sqrt{2}$.  This value is  valid 
for IceCube that observes only half of the sky.
For a neutrino telescope placed in the mediterranean sea
at a latitude of  approximately 36$^\circ$  one can  estimate
 $\langle a^{3/2} \rangle \simeq 1.11$, that is  20\% lower.
For a neutrino detector placed at the equator one should have
 $\langle a^{3/2} \rangle \simeq 1$.
The  dependence of the expected number of 
resolve sources  on  the detector  latitude
is an elementary   geometric  effect associated with
the reduced  field of  view of  a detector
close to the poles.
A  detector  that observes 
only a  fraction $1/n$ of the  celestial sphere
(with constant   effective area in this  portion of the sky) 
has  $\langle a^{3/2} \rangle = \sqrt{n}$.
The ``horizon''  of such a  detector
is $\sqrt{n}$ time  deeper that   the horizon of a detector
that  observes the entire sky with the same
angle integrated  exposure, and the number of  observed  sources
scales as the volume contained inside the horizon,
that is  $n^{-1} \times n^{3/2} = \sqrt{n}$ larger 
that  for an isotropic detector. 
Deeper observations of a smaller  fraction of the sky yield more
resolved sources.

The number of multiplets is also  proportional
to the total number of  events. This relation can be written
as:
\begin{eqnarray}
N_{\rm multi} 
& = & 
N_{\rm events} ~
\frac{1}{6 \, \sqrt{2} \;  R_0} 
\left [  \frac{A_0 \; t  \; \langle \varepsilon_{\nu \to \mu}\rangle \;  L_0 \; \eta}
{E_{\rm min}}\right ]^{1/2}
~\frac{\langle a^{3/2} \rangle ~\langle y^{3/2} \rangle
}{\zeta_\Phi}
\nonumber \\
& ~ & \nonumber \\
& = & 
2.9 \times 10^{-3} ~
~\sqrt{ \frac{L_0}
            { 10^{42} ~{\rm erg}/{\rm s}} }
~\sqrt{ \frac{A_0 ~ t}{{\rm Km}^2{\rm yr}} }
~\left (\frac{2.2}{\zeta_\Phi} \right )
 ~\langle y^{3/2} \rangle
~N_{\rm events}
\end{eqnarray}
This  can also  be rewritten as  a function
of the quantity  $K_{\nu_\mu} = E^2 ~\phi_{\nu_\mu} (E)$:
\begin{eqnarray}
N_{\rm multi} 
& = & 
0.47
\;\left (
\frac{K_{\nu_\mu}}{10^{-8} \;{\rm GeV}/{\rm cm}^2{\rm s}{\rm sr}}  \right )
~\sqrt{\frac{L_0}
             { 10^{42} \; {\rm erg}/{\rm s}} }
\; \left (\frac{A_0 ~  t}{ {\rm km}^2~{\rm  yr} } \right )^{3/2}
\;\left (\frac{2.2}{\zeta_\Phi} \right )
 \;\langle y^{3/2} \rangle
\end{eqnarray}

\begin{figure} [hbt]
\centerline{\psfig{figure=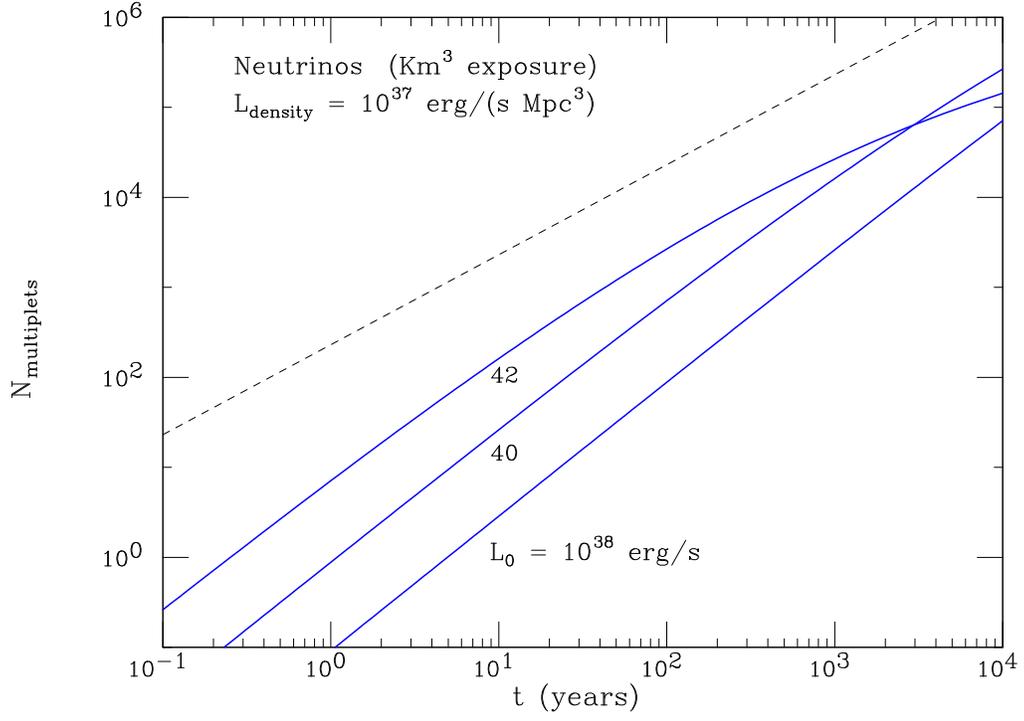,angle=90,width=13.5cm}}
\caption {\footnotesize 
Expected  number of  extragalactic  neutrino  sources  detectable in 
IceCube  with  a multiplicity  of up--going muons
$n \ge 2$ plotted as a function of detector  live time.
The  neutrino sources are all identical  and 
emit with power law  spectra with  slope  $\alpha = 2$
and a luminosity per  energy decade  $L_0$.
The power density of the ensemble of all
neutrino sources is
${\cal L} = 10^{37}$~erg/(s~Mpc$^3$).
The different curves are  calculated 
for different values  
$L_0 = 10^{38}$,
$10^{40}$ and
$10^{42}$ erg/s.
The dashed  line give  the inclusive  number of
detected up--going muons    that are produced by
extragalactic neutrinos.
For a  detector  places  in the Mediterranean  sea
at a latitude of $36^\circ$  the number of  sources
should be reduced by a factor 0.787.
\label{fig:nu_rate}  }
\end{figure}

For  completeness, figure~\ref{fig:nu_rate} shows  
a prediction of the evolution of the number
of clusters of  neutrino   induced muons  
produced by  extragalactic  neutrino sources
with increasing exposure.
The  number of multiplets is expected to grow approximately $\propto (A \, t)^{3/2}$ 
even for exposures   of  hundreds of square kilometer year.
This  is a consequence of the fact that this functional
form continues  until   the
detector sensitivity  is sufficient to  see
as   muon multiplets sources  located at approximately a Hubble distance,
and this   does require very large   exposures, unless the luminosity of the source
is extraordinarily large.

\section{Outlook}
The key problems   about  the  nature and properties of 
the extragalactic cosmic   rays  sources  remain open,
even if some  of the answers seems to be   tantalizingly close.
One of the most urgent   problems is the determination of  the 
region of   source distance  and   particle  energy where   astronomy 
with charged cosmic rays  is possible.
The   results of Auger  on the correlation of the arrival direction of
the highest energy CR with the positions  of close  AGN    suggest
that astronomy with CR   is possible  with a threshold
of    $\sim 6 \times 10^{19}$~eV,  and   a volume 
of radius   $R \simeq 75$~Mpc. These are very 
encouraging results,   but  they  need to be confirmed  with  more statistics.
If a  significant part of the anisotropy effect  observed by Auger is due
to the contribution of a single source at  only 3.5~pc    the observation of  more  distant sources
could be significanly more  problematic.

Assuming that  the linear  propagation of the highest energy cosmic  rays  is  a reasonable approximation
and that  most of the particles  are protons, the clustering (and lack of clustering)  of the
highest energy events  determinee to within a   factor of 300 or so the    typical luminosity 
of  a extragalactic CR sources
(in the   energy decade above $E \simeq 6 \times 10^{19}$~eV) as
$L_0 \in [10^{39},10^{41.5}]$~erg/s. This is not inconsistent with the estimated  luminosity
of Cen~A  that is of order few times  $10^{39}$~erg/s. 
In the same  energy region the power density of the   ensemble of the  CR sources if  of order
of $1--2 \times 10^{36}$~erg/(s~Mpc$^3$).

The neutrino  observations, that at present  limit the 
power  density  of the neutrino sources  to
a  maximum value of order ${\cal L}_\nu \lesssim 10^{37}$~erg/(s~Mpc$^3$)
offer  already a  quite  significant constraint, since the neutrino limit
refer to  the power density    per   energy decade in  
at much lower  energy,  and therefore suggest that the  spectrum 
of the CR  emission  cannot extend to low energy 
with a too  steep spectrum.
The inclusive extragalactic neutrino flux   is mostly  generated
by  distant, very faint sources, and therefore
only a small fraction (probably less than $10^{-3}$)   can be 
resolved in the contribution   of individual sources by detectors of 
the IceCube type.
If an extragalactic   neutrino component is not  too   far from  
the existing upper limits, the sensitivity of  the Km$^3$ neutrino telescopes 
could be insufficient to detect  more than a handful of  extragalactic sources.

\vspace {0.4 cm}
\noindent{\bf Acknowledgments}  I would like  to acknowledge discussions with 
Steve Barwick, Maurizio Lusignoli,  Andrea Silvestri and Todor Stanev.

\appendix

\section{Detector Acceptance for EAS Arrays}
\label{sec:app_eas}
The geometry of  an Extensive  Air Shower Array  (EAS) 
 is   sufficiently simple
that it is possible to have a   reasonably accurate 
description of  its acceptance  with simple  analytic  expressions.
In this type of detection method, cosmic  ray  showers are 
analysed if their  zenith angle  is  smaller than a 
maximum    value $\theta_{\rm max}$, and if  the shower  axis
falls inside a  predetermined  region of area $A_0$.
The  effective area  for  the detection of  showers   coming from a  direction
with zenith angle $\theta$  is therefore  
$A(\theta) \simeq A_0 \; \cos \theta$.
The angle  integrated area is:
\begin{equation}
A_\Omega 
=  A_0 ~  2 \pi \; \int_{\cos\theta_{\rm max}}^1
d\cos \theta ~\cos \theta
=  A_0 ~  \pi  \, 
\left [1
-  (\cos \theta_{\rm max})^2 \right ]
%\simeq  
\end{equation}
The Auger  detector  has   used the cut $\theta_{\max} = 60^\circ$, and 
therefore its  integrated  exposure is approximately:
\begin{equation}
A_\Omega \simeq   A_0 ~ \left (\frac{3 \, \pi }{4}  \right )
\end{equation}

\begin{figure} [hbt]
\centerline{\psfig{figure=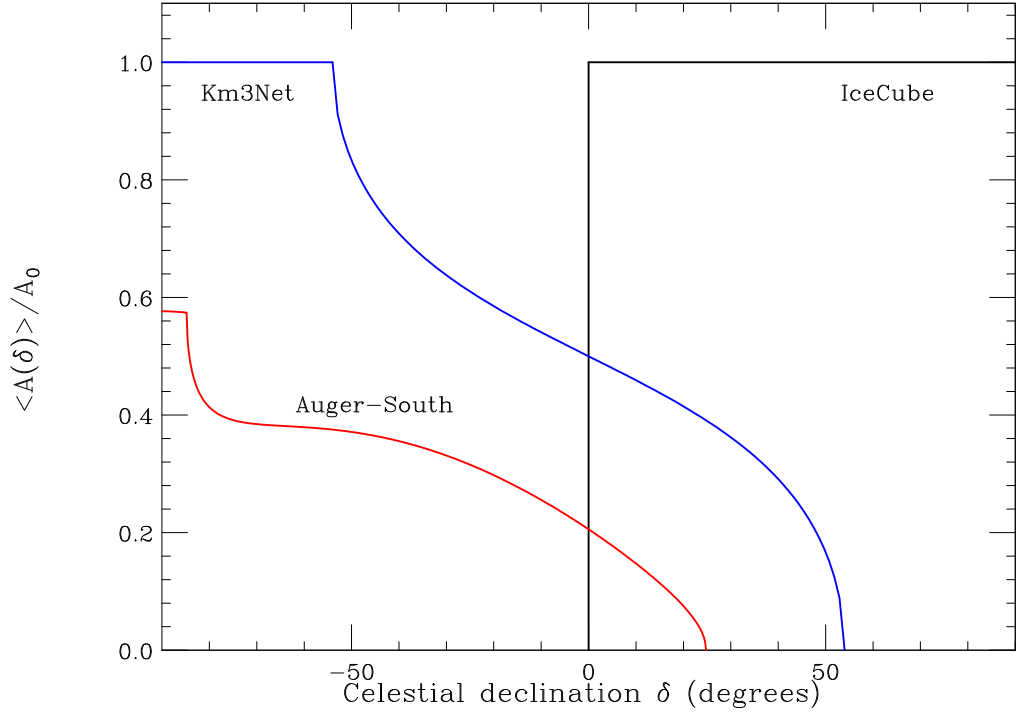,angle=90,width=13.5cm}}
\caption {\footnotesize 
Plot of the  detector  effective  area  $A(\Omega)$ 
as a function of the  celestial  declination $\delta$.
In  the case of the Auger--South detector, the  effective area
is divide   by the surface  area of the array, 
the detector  latitude  is  $\varphi_{\rm lat} = -24.8^\circ$  
and the projected  area  is  averaged
for the time  when a   source  has  zenith angle $\theta < 60^\circ$.
For the  neutrino  telescopes, the   detector projected area 
in   the direction $\theta$  is approximated as constant, and 
$\nu$ detection
is  considered  possible   for all  zenith angles $\theta > 90^\circ$.
\label{fig:expos}  }
\end{figure}

The interesting problem  is  to compute  the detector
acceptance  as  a function of the  celestial  direction.
This can  be done  taking into account the trajectories
in  the  detector system  of points with different  
 coordinates on the celestial sphere.
All points with the same  celestial declination $\delta$  make the same
trajectory in  the sky  with  zenith angle:
\begin{equation}
\cos \theta (h, \delta, \varphi_{\rm lat})
= \cos \delta \; \cos \varphi_{\rm lat} \; \cos h +
\sin \delta \; \sin \varphi_{\rm lat} 
\end{equation}
($h$  is the hour angle).
Averaging  over  one  sidereal  day and taking into account only 
of the  time when a  celestial  direction  is  sufficiently
high above the horizon, one can   compute  the ``average''  
effective  area for  points   with   declination   $\delta$ as:
\begin{equation}
\langle
A(\delta) \rangle =
A_0 ~\langle
\cos \theta (\delta, \varphi_{\rm lat} )
\rangle =
A_0~\int_0^{2 \pi} \frac{dh}{2 \pi} 
~\cos \theta (h, \delta, \varphi_{\rm lat})
~\Theta \left [\cos \theta - \cos \theta_{\rm max} \right] 
\end{equation}
The averaging over a sidereal  day is  a good approximation 
for  long periods  of  data  taking.
The  function 
$\langle \cos \theta (\delta, \varphi_{\rm lat}  \rangle$
for the latitude of the Auger  detector ($\varphi_{\rm lat} \simeq  24.8^\circ$)
is  shown in  fig.~\ref{fig:expos}.
Integrating  the acceptance  over all    the celestial  sphere
one must obtain the same  result  as integrating over  the detector
angular coordinates,  and therefore  one  obtains  
(for  any detector  latitude) the relation:
\begin{equation}
\int_{-\pi/2}^{\pi/2}
d \delta \cos  \delta  ~
\langle \cos \theta (\delta, \varphi_{\rm lat} ) \rangle
= \frac{1}{2} 
\, \left [1 -  (\cos \theta_{\rm max})^2 \right ]
\end{equation}

\section{Neutrino  detection}
\label{sec:app_nu}
  The most promising  technique for 
  the detection  of  point--like astrophysical  neutrinos  sources
  is  the   observations  of $\nu$--induced  muons. 
  These  muons are produced in  the charged current
  interactions of $\nu_\mu$ and  $\overline{\nu}_\mu$ in the matter below
  the detector, and  are  approximately collinear
  with the   parent neutrino  direction, and 
therefore  excellent tools  for the imaging of neutrino sources.

  The  $\nu$--induced muon   signal is  detectable only
  eliminating the large background  of atmospheric  muons
  due to the decay of charged pions  and kaons
produced  in  the  hadronic  showers of 
cosmic rays interacting in the Earth's atmosphere.
This   can be achieved  limiting 
the neutrino  observations to  up--going  trajectories.
With this limitation, a 
 neutrino  telescope  placed at the South Pole can only observe
the northern  celestial  hemisphere.

The   signal  of astrophysical  neutrinos remains partially hidden
by the   background 
of  atmospheric  neutrinos   that  has the same
origin of the muon  background
(the weak decays   of hadronic  produced by 
cosmic  ray  showers).
The astrophysical neutrino signal can however 
be  disentangled  from  this background  because
of its different  (harder)  energy  spectrum.
Point sources  can  be identified as  clusters
of  events with approximately the same  direction.

In this  work, following \cite{Lipari:2006uw}, 
we  will consider a simplified  description   of  neutrino  detection
that is  sufficient for  our discussion.
We will  assume that the neutrino  energy
can be reasonably well  described 
in the relevant energy range 
($10^{12} \lesssim E_\nu \lesssim 10^{16}$~eV) as  a simple
power law of slope  $\alpha$.
The flux of muons   generated by   neutrinos with spectrum $\phi_\nu (E_\nu)$ 
can be calculated as:
 \begin{eqnarray}
 & & \Phi_\mu (E^{\rm min}_\mu  ) =
 \int dE_\nu  ~\frac{\phi(E_\nu)}{6} 
  \left [ P_{\rm surv}^{\nu_\mu} (E_\nu, \theta)~
  Y_{\nu_\mu \to \mu^-} (E_\nu, E^{\rm min}_\mu ) 
\right . \nonumber \\
 &  &  \left . ~~~~~~~~~~~~~~~~~~~~~~~~~~~~~~~~~~~~ + 
 P_{\rm surv}^{\overline{\nu}_\mu} (E_\nu, \theta)~
 Y_{\overline{\nu}_\mu \to \mu^+} (E_\nu, E^{\rm min}_\mu )
 \right ]
 \label{eq:nu_det}
 \end{eqnarray}
 In this expression  $\phi_\nu(E_\nu)$ is the
 neutrino flux summed over all $\nu$  types, and the factor
 1/6  follows  from the assumption   that   flavor  oscillations
 mix the neutrinos,  so that one has  approximately
 equal amounts  for each type
 ($\nu_e$,  
  $\overline{\nu}_e$
 $\nu_\mu$,
  $\overline{\nu}_\mu$
 $\nu_\tau$ and
  $\overline{\nu}_\tau$).
For a critical  discussion of this    result see for  example
\cite{Lipari:2007su}.
 The function 
$P_{\rm surv}^{\nu_\mu} (E_\nu)$ 
($P_{\rm surv}^{\overline{\nu}_\mu} (E_\nu)$)
is     the probability
 that a  $\nu_\mu$  ($\overline{\nu}_\mu$) that arrives  
at the  detector with zenith angle $\theta$,
survives  the  crossing the Earth without  interacting, 
and   $Y_{\nu_\mu \to \mu^-}(E_\nu, E_{\mu ,{\rm min}})$ 
 ($Y_{\overline{\nu}_\mu \to \mu^+}(E_\nu, E_{\mu ,{\rm min}})$) is the 
muon  yield, that is the 
number  of muons   
with  $E \ge E_{\mu, {\rm min}}$  generated  by a $\nu_\mu$
 ($\overline{\nu}_\mu$)  of energy $E_\nu$.

\begin{figure}[ht]
\centerline{\psfig{figure=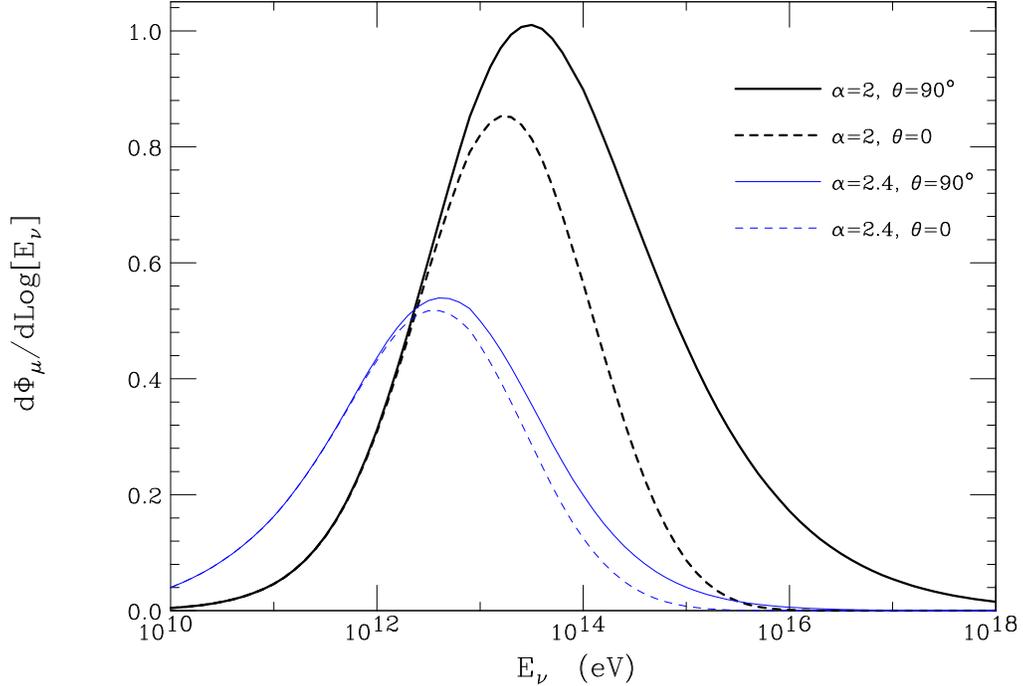,angle=90,width=13.5cm}}
    \caption{\footnotesize 
Neutrino  energy response for  the up--going muon signal.
The area under the curves is  proportional to the
signal of $\nu$ induced muons 
(with threshold $E_\mu = 1$~GeV)  assuming a  power law neutrino  flux
with slope $\alpha$. The thick (thin)  curves are calculated
for  a slope $\alpha = 2$ ($\alpha = 2.4$); the solid
(dashed) curves  are calculated  for
a nadir  angle $\theta = 90^\circ$   ($\theta = 0^\circ$)
and negligible  (maximum)  $\nu$ absorption in the Earth.
The relative normalization of the curves  is correct if the spectra
have the same   integrated flux above 1~TeV.
The $\nu$ induced  signal   is  roughly proportional to
 the integrated  neutrino flux  in the approximate 
energy range  $[10^{12}, 10^{16}]$~eV
in case of a hard spectrum, or
$[10^{11}, 10^{15}]$~eV
for a  soft  spectrum.
 \label{fig:nu_response}  }
 \end{figure}

\begin{figure}[ht]
\centerline{\psfig{figure=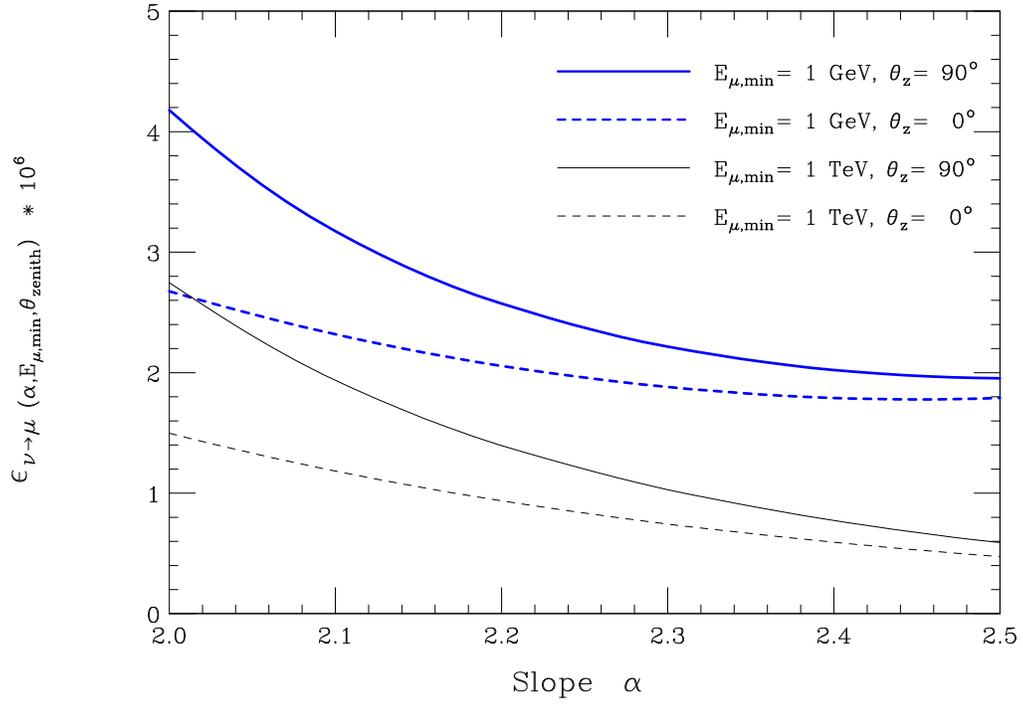,angle=90,width=13.5cm}}
    \caption{\footnotesize 
 Neutrino to muon  conversion efficiency.
 The different  curves  correspond  to different choices for the
 muon  detection threshold and different celestial declination of a source
 for a  neutrino  detector at the South Pole.
 \label{fig:muon_eps}  }
 \end{figure}

The integrand of equation (\ref{eq:nu_det})  is  shown in  fig.~\ref{fig:nu_response}, 
for two   examples of the slope $\alpha$ and
on the two  extreme  cases   (minimum and maximum) 
of the effects of absorption  in the Earth.
Inspection of  fig.~\ref{fig:nu_response}  shows
that   for a neutrino spectrum   of slope not much larger
than $\alpha \simeq 2$ the muon signal is   generated
by neutrinos  in the energy  range $[10^{12},10^{16}]$~eV.
It is  therefore  useful, at least for a qualitative understanding
to consider the     flux the neutrino induced muons
as   proportional to the integral flux    of neutrino  above
a (conventional  but   roughly appropriate)  energy threshold
of $E_\nu^{\rm min} = 10^{12}$~eV, introducing
an energy averaged  neutrino to muon  conversion efficiency $\langle \varepsilon_{\nu \to \mu}\rangle$
that can be calculated performing the ratio
between the muon and the neutrino fluxes:
 \begin{equation}
\langle \varepsilon_{\nu \to \mu}\rangle
 =  \frac{\Phi_{\mu \uparrow} (E^{\rm min}_\mu )} { \Phi_\nu (E_\nu \ge 1~{\rm TeV}) }
 \end{equation}
The conversion efficiency $\langle \varepsilon_{\nu \to \mu}\rangle$  depends    on
(i) the muon energy threshold,
(ii) the slope of the neutrino spectrum (being  higher  for  flatter  spectra),
and because of the effects of $\nu$ absorption in the Earth,  (iii) on the
neutrino  zenith  angle.
This is illustrated   in fig.~\ref{fig:muon_eps})
that  shows  some examples of   $\langle \varepsilon_{\nu \to \mu}\rangle$.  
In the following,  we will  neglect the zenith  angle  dependence
of  $\langle \varepsilon_{\nu \to \mu}\rangle$, considering that this is  a  reasonable approximation
given the size of the astrophysical  uncertainties.

For a  detector  located at the South Pole,  the northern  celestial
hemisphere remains  constantly  below the  horizon  and can
observed  continuosly,  while the southern hemisphere  in
unobservable.  The effective area will  be simply modeled as:
 \begin{equation}
 A(\Omega) \simeq \cases {
A_0 ~\langle \varepsilon_{\nu \to \mu}\rangle   & for $\delta > 0$ \cr
0   & for $\delta \le  0 $
}
\end{equation}
with $A_0 \simeq 1$~Km$^2$.
The angle integrated effective area is  $A_\Omega = 2 \, \pi \; A_0 ~\langle \varepsilon_{\nu \to \mu}\rangle$.
For  first order  numerical  estimates in  the following
we will use $\langle \varepsilon_{\nu \to \mu}\rangle = 3 \times 10^{-6}$, that is the correct value
for a low  muon  energy threshold and a slope $\alpha \simeq 2.1$.
For  a flatter  spectra with $\alpha \simeq 2.5$, the conversion is 
smaller:  $\langle \varepsilon_{\nu \to \mu}\rangle \simeq 2 \times 10^{-6}$.

%%%%%%%%%%%%%%%%%%%%%%%%%%%%%%%%%%%%%%%%%%%%%%%%%%%%%%%%%%%%%%%%%%%%%

%%%%%%%%%%%%%%%%%%%%%%%%%%%%%%%%%%%%%%%%%%%%%%%%%%%%%%%%%%%%%%%%%%%%%

%%%%%%%%%%%%%%%%%%%%%%%%%%%%%%%%%%%%%%%%%%%%%%%%%%%%%%%%%%%%%%%%%%%%%

%%%%%%%%%%%%%%%%%%%%%%%%%%%%%%%%%%%%%%%%%%%%%%%%%%%%%%%%%%%%%%%%%%%%%

\clearpage

%%%%%%%%%%%%%%%%%%%%%%%%%%%%%%%%%%%%%%%%%%%%%%%%%%%%%%%%%%%%%%%%%%%%%

%%%%%%%%%%%%%%%%%%%%%%%%%%%%%%%%%%%%%%%%%%%%%%%%%%%%%%%%%%%%%%%%%%%%%

%%%%%%%%%%%%%%%%%%%%%%%%%%%%%%%%%%%%%%%%%%%%%%%%%%%%%%%%%%%%%%%%%%%%%

%%%%%%%%%%%%%%%%%%%%%%%%%%%%%%%%%%%%%%%%%%%%%%%%%%%%%%%%%%%%%%%%%%%%%

%%%%%%%%%%%%%%%%%%%%%%%%%%%%%%%%%%%%%%%%%%%%%%%%%%%%%%%%%%%%%%%%%%%%%

%%%%%%%%%%%%%%%%%%%%%%%%%%%%%%%%%%%%%%%%%%%%%%%%%%%%%%%%%%%%%%%%%%%%%

%%%%%%%%%%%%%%%%%%%%%%%%%%%%%%%%%%%%%%%%%%%%%%%%%%%%%%%%%%%%%%%%%%%%%

%%%%%%%%%%%%%%%%%%%%%%%%%%%%%%%%%%%%%%%%%%%%%%%%%%%%%%%%%%%%%%%%%%%%%

\clearpage

%%%%%%%%%%%%%%%%%%%%%%%%%%%%%%%%%%%%%%%%%%%%%%%%%%%%%%%%%%%%%%%%%%%%%

%%%%%%%%%%%%%%%%%%%%%%%%%%%%%%%%%%%%%%%%%%%%%%%%%%%%%%%%%%%%%%%%%%%%%

%%%%%%%%%%%%%%%%%%%%%%%%%%%%%%%%%%%%%%%%%%%%%%%%%%%%%%%%%%%%%%%%%%%%%

%%%%%%%%%%%%%%%%%%%%%%%%%%%%%%%%%%%%%%%%%%%%%%%%%%%%%%%%%%%%%%%%%%%%%

%%%%%%%%%%%%%%%%%%%%%%%%%%%%%%%%%%%%%%%%%%%%%%%%%%%%%%%%%%%%%%%%%%%%%

%%%%%%%%%%%%%%%%%%%%%%%%%%%%%%%%%%%%%%%%%%%%%%%%%%%%%%%%%%%%%%%%%%%%%

%%%%%%%%%%%%%%%%%%%%%%%%%%%%%%%%%%%%%%%%%%%%%%%%%%%%%%%%%%%%%%%%%%%%%

\end{document}